\documentclass[a4paper,times]{cjeart}

\usepackage{multirow}
\usepackage{makecell}
\usepackage{subfig}
\usepackage{amsthm} 

\graphicspath{{figures/}}
\usepackage{algorithm,algorithmic}

\volumeyear{2025}
\volumenumber{04}
\issuenumber{04}
\journalname{Chinese Journal of Electronics}
\startpage{1}
\DOI{123.123.123.123}
\journaltype{Research Article}

\title[Fundamental Models and Signal Processing for Movable Antenna-Enhanced Wireless Communications and Sensing]{Fundamental Models and Signal Processing for Movable Antenna-Enhanced Wireless Communications and Sensing}

\author{%
Zhenyu Xiao\affilnums{1}, 
Xiangyu Pi\affilnums{1}, 
Songqi Cao\affilnums{1}, 
Lipeng Zhu\affilnums{2}, 
Zhen Gao\affilnums{3}, 
Xiang-Gen Xia\affilnums{4}, and \\
Rui Zhang\affilnums{2}
}

\affiliation{%
\affilnum{1}School of Electronic and Information Engineering, Beihang University, Beijing 100191, China\\
\affilnum{2}Department of Electrical and Computer Engineering, National University of Singapore, Singapore 117583, Singapore \\
\affilnum{3}Schoolof Information and Electronics, Beijing
Institute of Technology, Beijing 100081, China \\
\affilnum{4}Department of Electrical and Computer Engineering, University of Delaware, Newark, DE 19716, USA
}

\corrauth{Lipeng Zhu}
\email{zhulp@nus.edu.sg}

\abstract{%
Movable antenna (MA) has been recognized as a promising technology for performance enhancement in wireless communication and sensing systems by exploiting the spatial degrees of freedom (DoFs) in flexible antenna movement. However, the integration of MAs into next-generation wireless networks still faces design challenges due to the  paradigm shift from conventional fixed-position antennas (FPAs) to MAs, which motivates this paper to provide a comprehensive overview of the models, scenarios, and signal processing techniques for MA-enhanced wireless networks. First, we introduce several efficient methods to realize flexible antenna movement. Next, channel models based on field response and spatial correlation are presented to characterize the channel variations with respect to MA movement.  Then, we discuss the advantages and challenges of applying MAs to typical application scenarios of wireless communications and sensing. Moreover, we show the signal processing techniques for MA-enhanced communication and sensing systems, including channel acquisition and antenna position optimization. Finally, we highlight promising research directions to inspire future investigations.
}


\begin{document}

\maketitle

\section{Introduction}
\label{s1}
The evolution of mobile communication networks  to the sixth generation (6G)   renders unprecedented demands for ultra-high data rate, hyper reliable and low latency, high precise positioning, global coverage, and ubiquitous connectivity~\cite{6G1,6G2}, which are essential for future applications such as Internet of Everything (IoE), automated driving, augmented reality (AR), and virtual reality (VR)~\cite{6G3,6G4}. To improve the spectral efficiency,  
multiple-input multiple-output (MIMO) has been recognized as a revolutionary technology in mobile networks by shifting from single-antenna to multi-antenna systems~\cite{MIMO1}. The increasing number of antennas in MIMO systems not only attains  higher array/beamforming gains to compensate for signal propagation loss, but also enhances the spatial multiplexing gains for serving multiple users simultaneously over the same time/frequency resource block, which drives the development of MIMO technologies towards more advanced forms, such as massive MIMO  and extremely large-scale MIMO (XL-MIMO), to support the emerging applications in  6G~\cite{mMIMO,XLMIMO}.  However, the excessively large number of antennas and their associate radio-frequency (RF) chains  inevitably introduce severe challenges in terms of higher hardware cost, energy consumption, and signal processing overhead.

For this reason, significant research efforts have
been devoted to improving energy and spectral efficiency in MIMO systems. Specifically, thinned arrays and sparse arrays have been extensively explored to reduce the number of active antennas while maintaining the system's performance~\cite{Thinned, Sparse}. In addition, beamspace MIMO enables cost-efficient implementation by utilizing lens antenna arrays, which helps minimize the number of RF chains and phase shifters at the transceiver~\cite{beamMIMO1,beamMIMO2}. Intelligent reflecting surface (IRS), which is fabricated at a lower cost through the integration of large numbers of semi-passive and tunable reflective elements, has been widely investigated due to its superior capabilities of reconfiguring wireless propagation
environments and enhancing signal coverage~\cite{RIS1,RIS2}.   Furthermore, dynamic metasurface antennas (DMAs) and reconfigurable holographic surfaces (RHSs) employ metamaterial elements as active radiating components~\cite{DMA,RHA}. These components are arranged in high density to attain high aperture gains and beamforming performance, eliminating the need for conventional high-cost phase shifters.

However,  the aforementioned implementations of both active and passive arrays are based on fixed-position antennas (FPAs), whose geometries are fixed once manufactured.  The limitation of FPAs stems from their insufficient spatial degrees of freedom (DoFs), which severely constrains their adaptability in dynamic wireless environments. This spatial rigidity renders FPA systems highly susceptible to performance degradation caused by random channel fading, transmission blockages, and severe interference, as they cannot physically reconfigure antenna positions to track mobile users or mitigate interference. In sensing applications, this static nature results in a fixed perspective with a limited angular resolution and possible blind spots, drastically reducing their ability to accurately locate and track moving targets or adapt to changes in the surrounding radio environment.

To overcome these fundamental limitations, movable antenna (MA) has emerged as a promising technique, which fully exploits the spatial DoFs to enhance wireless communication and sensing performance by flexibly adjusting antenna position in the continuous spatial regions at the transmitter (Tx)/receiver (Rx)~\cite{Zhu2024mag,MASISO_NB,MIMO_FFC}, which is sometimes also referred to as fluid antenna~\cite{zhu2024his}. Specifically, MAs can dynamically change their placements to escape from the positions experiencing deep fading channels and/or severe interference, thereby effectively improving communication performance. Moreover, different from
conventional FPA arrays,  the geometry of an MA array can be configured to  enable more flexible beamforming, which is essential for adapting to dynamic user distributions and mitigating co-channel interference of wireless networks, especially in dense deployment scenarios (e.g., urban macrocells and indoor hotspots)~\cite{Zhu2025Tut,shao20256DMA}.

\begin{figure*}[!htb]
	\centering	
	\subfloat[\label{fig:a}IoT]{\includegraphics[width=.3\linewidth]{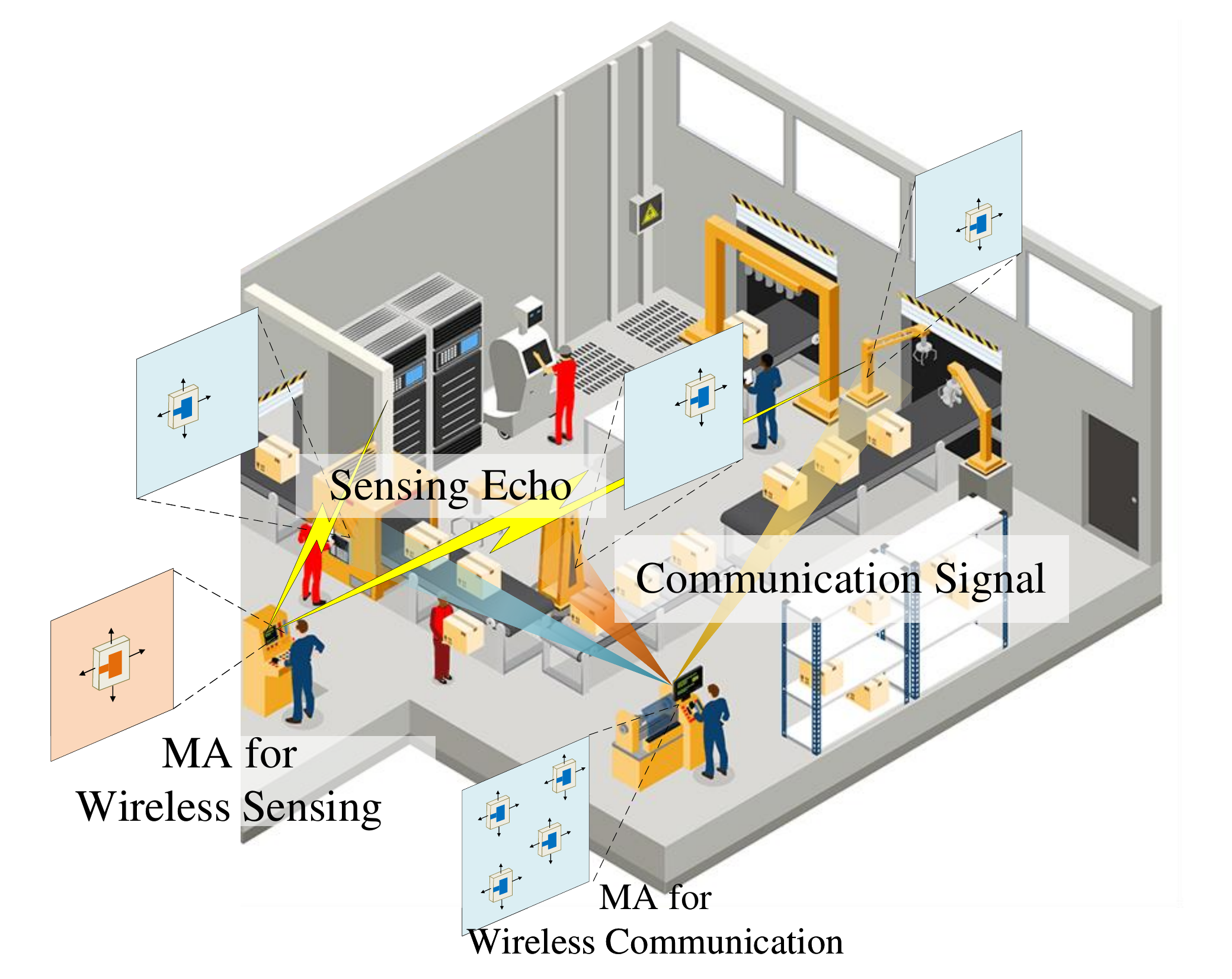}}%
	\subfloat[\label{fig:b}Mobile 
	network]{\includegraphics[width=.3\linewidth]{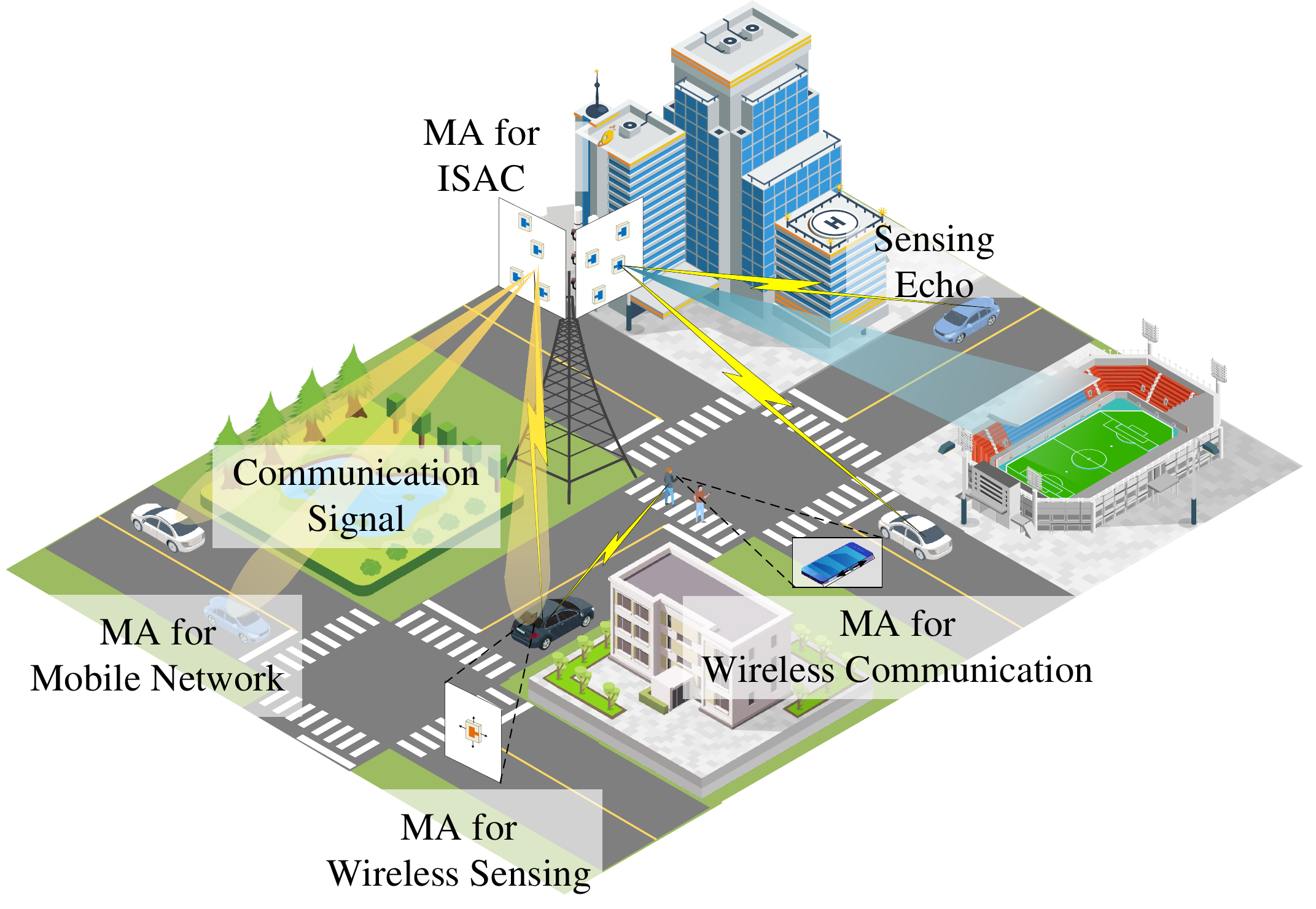}}
	\subfloat[\label{fig:c}NTN]{\includegraphics[width=.3\linewidth]{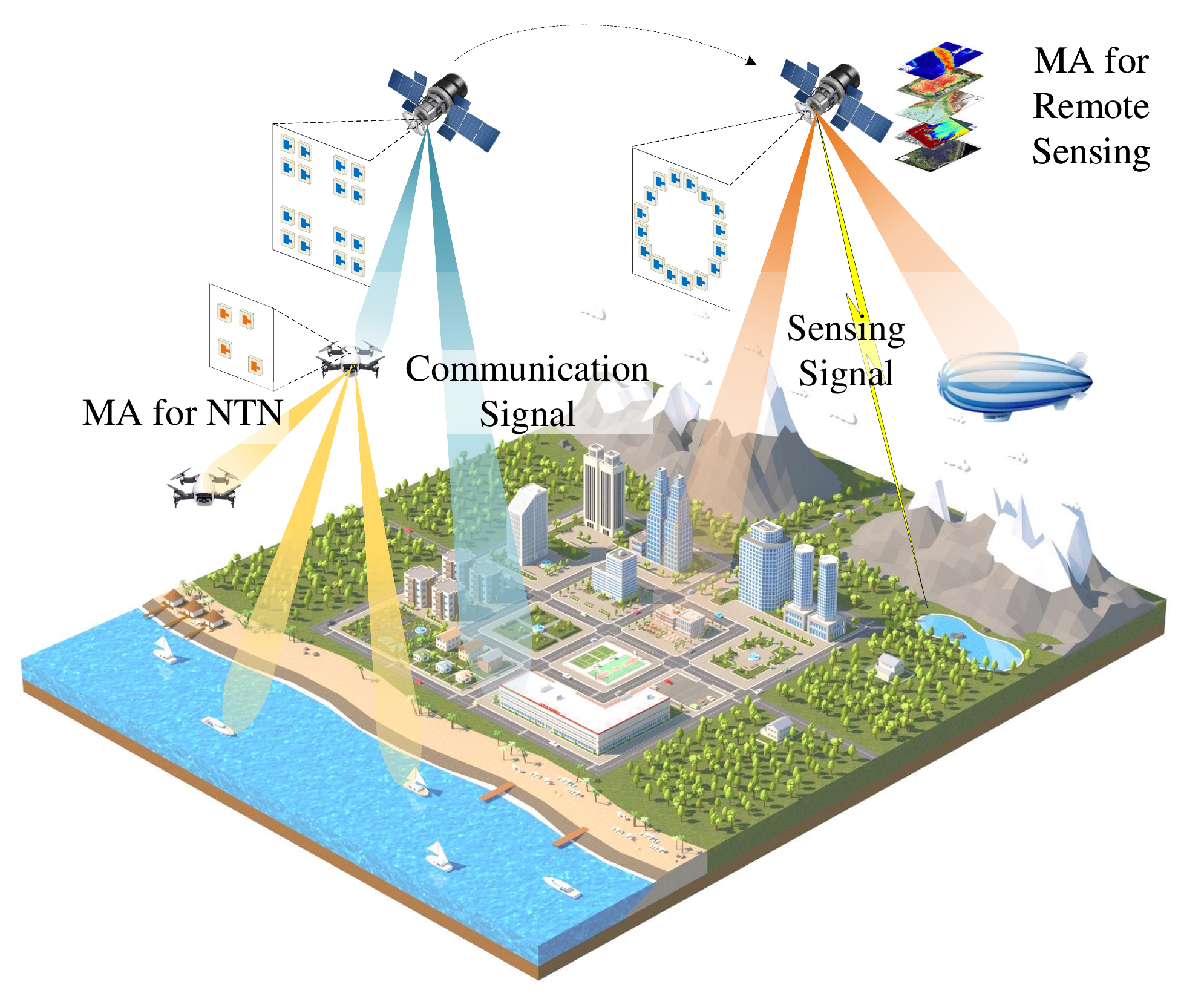}}
	\caption{Typical applications for MA-aided wireless networks.}
	\label{fig:1}
\end{figure*}
Given the above benefits, MAs are promising to be applied to various wireless communication and sensing systems, with the typical application scenarios shown in Fig.~\ref{fig:1}. In Fig.~\ref{fig:1}\subref{fig:a}, MA-enhanced  Internet of Things (IoT) is illustrated, where MAs  are deployed on a massive number of machine-type communication (MTC) devices. Therefore, the spatial variation of the wireless channel can be fully exploited by antenna position optimization of MAs on MTC devices, thereby effectively addressing the issue of time-frequency resource constraints. 
In Fig.~\ref{fig:1}\subref{fig:b}, MA-enhanced mobile network is illustrated, where MAs
can be deployed on base stations (BSs) and vehicles. Due to the mobility of users and scatterers, the ensuing time-varying channels pose a huge challenge to the reliability and stability of wireless communications and sensing. In light of this challenge, MAs introduce an additional spatial DoF by enabling dynamic reconfiguration of antenna placement. This capability significantly improves channel quality for mobile users. Such flexibility promotes more efficient use of temporal, spectral, and power resources, thereby substantially enhancing overall network performance.
In practical implementations, the time interval of antenna movement  can be flexibly adjusted to accommodate either instantaneous or statistical channel changes. This enables a favorable balance between system performance and the movement overhead. For instance, in scenarios involving low-mobility users with slowly varying channels, antenna adjustments can be performed based on real-time channel state information (CSI). Conversely, for high-mobility users experiencing rapid channel variations, antenna configurations may be optimized using long-term statistical CSI, thereby avoiding the need for frequent repositioning in practical deployments.
In Fig.~\ref{fig:1}\subref{fig:c}, MA-enhanced non-terrestrial network (NTN) is illustrated, where MAs can be deployed on satellites and aircraft.
In conventional NTN, satellites are generally equipped with large-scale antenna arrays for synthesizing narrow beams to improve the angular resolution, and for enhancing beam gains to compensate for the high path loss. However, the satellite-mounted FPA arrays generally face challenges in severe coupling effects and heat dissipation in dense
arrays, as well as undesirable sidelobes and interference
leakage in sparse arrays. In addition, the FPA arrays fail to utilize the full DoF in beamforming to keep up with the fast-varying channel conditions and system requirements. In contrast, the geometry of an MA array can be reconfigured such that more flexible beam patterns can be obtained, where the mainlobe 
can be narrowed while suppressing the sidelobes. Moreover, the flexible beamforming can be achieved by jointly optimizing the positions and weights of MAs, thus guaranteeing the quality of service (QoS) of the coverage area while suppressing the power leakage of the interference area.  

The opportunities and challenges of MA-aided wireless networks were preliminarily investigated in~\cite{Zhu2024mag}, which revealed the superiority of MA systems over FPA systems in terms of received signal power improvement, interference mitigation, flexible beamforming, and spatial multiplexing enhancement. In~\cite{Zhu2025Tut}, a tutorial on MA-aided wireless networks was
conducted, where the historical development, design
issues, MA fundamentals, and prototypes were presented. Lately, the authors in~\cite{Shao2025Tut} extended MA to six-dimensional MA (6DMA) by exploiting both three-dimensional (3D) positions and 3D rotations, and discussed the challenges and potential research directions. Moreover,  the authors in~\cite{wang2025Tut} reviewed the flexible MIMO technology, which was classified into designs based on flexible deployment characteristics, flexible geometry characteristics, and flexible real-time
modifications.

Different from the aforementioned works~\cite{Zhu2024mag,Zhu2025Tut,Shao2025Tut,wang2025Tut}, we focus on the fundamental models and advanced  signal processing techniques for  MA-enhanced wireless communication and sensing  systems. 
In this paper, considering the unique characteristics
of MA-aided wireless networks, we first introduce the
antenna movement and channel models in~\ref{s2}. Then, the performance advantages of MAs in typical application scenarios for wireless communications and sensing are presented in~\ref{s3}. Next, we provide a comprehensive overview and comparison of the state-of-the-art signal processing techniques for both antenna position optimization and channel acquisition in~\ref{s4}, highlighting their advantages and development trend. Finally, we summarize open problems worthy of future research in this fertile area in Section~\ref{s5}.

\textit{Notations}: $a$, $\mathbf{a}$, $\mathbf{A}$, and $\mathcal{A}$ denote a scalar, a vector, a matrix, and a set, respectively. $\mathbb R$ and $\mathbb C$ represent the sets of real and complex numbers, respectively.  
$(\cdot)^{\rm{T}}$ and $(\cdot)^{\rm{H}}$ denote transpose and conjugate transpose, respectively. 
$[\mathbf{a}]_i$ and $[\mathbf{A}]_{i,j}$ denote the $i$-th entry of vector $\mathbf{a}$ and the entry in the $i$-th row and $j$-th column of
matrix $\mathbf{A}$, respectively. $||\cdot||_0$ and $||\cdot||_2$ denote the $l_0$-norm and $l_2$-norm, respectively.   In addition, $\otimes$ denotes the Kronecker product
$\mathcal{CN}(0,\sigma^2)$ represents the circularly symmetric
complex Gaussian (CSCG) distribution with mean zero and variance $\sigma^2$. 
\section{Fundamental Models}
\label{s2}
In this section, the state-of-the-art hardware implementations for realizing antenna movement are introduced and compared in terms of hardware complexity and movement performance. Then, the channel models are presented to characterize the channel variations with respect to antenna movement.
\begin{figure*}[t]
	\centering
	\includegraphics[width=\linewidth]{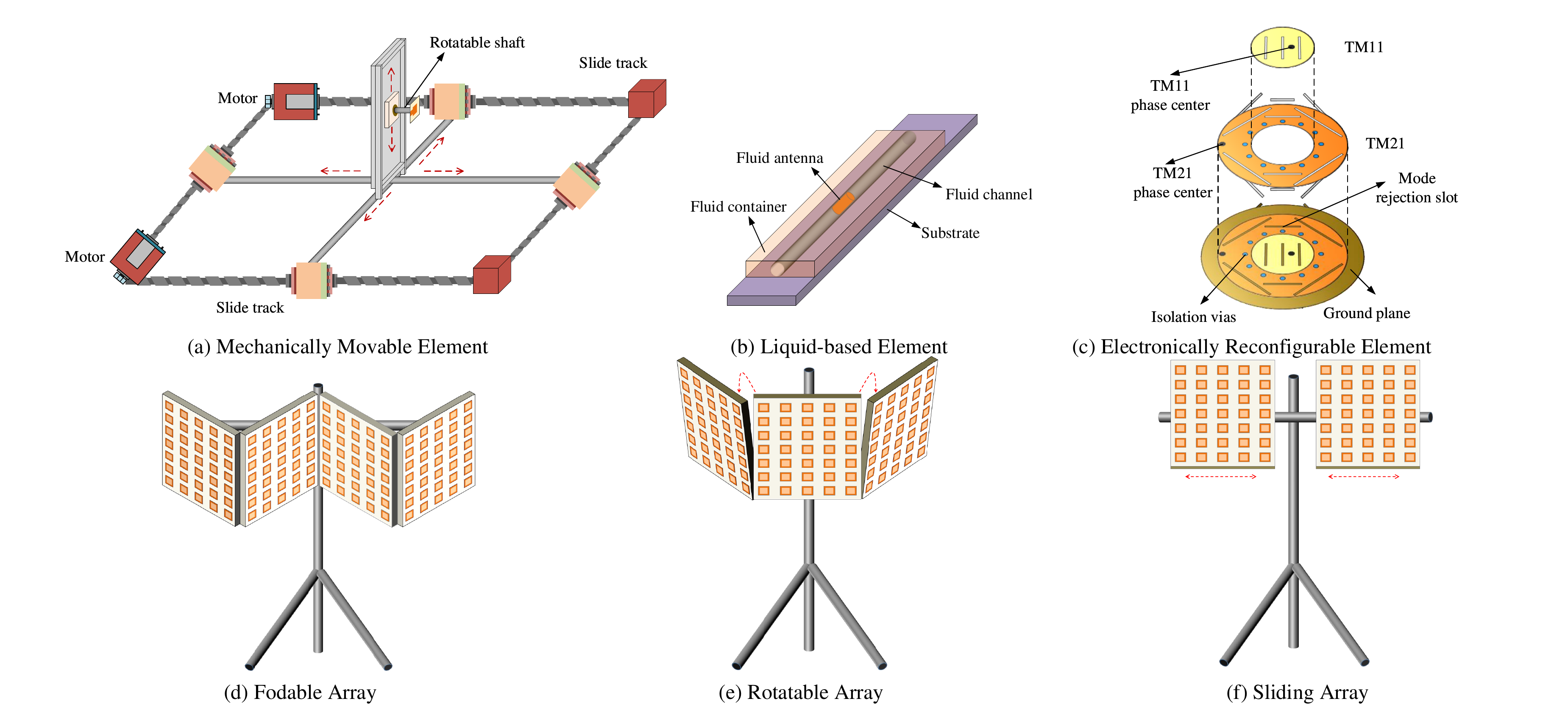}
	\caption{Illustration of typical antenna movement architecture.}
	\label{fig:MA_arch}
\end{figure*}
\subsection{Antenna Movement Architecture}
\label{s2-1}
To accommodate different system requirements and operating environments, the  architectures for implementing MAs have been widely investigated, which have been summarized and compared in Table~\ref{tab:arch}.
\begin{table*}[h]
	\renewcommand{\arraystretch}{1.3}
	\caption{Comparison of architectures for implementing MAs.}\label{tab:arch}
	\footnotesize
	\begin{center}
		\begin{tabular}{|c|c|c|c|c|c|}
			\hline
			\textbf{Architecture}
			&\textbf{Implementation} &\textbf{Movement Mode}          &\textbf{Response Time} &\textbf{Positioning Accuracy}     &\textbf{Ref.}\\
			\hline
			\multirow{2}*{\makecell{Mechanically\\movable\\element}}& Motor    &3D position                             	& \makecell{Millisecond to second order} & Micrometer order  &\cite{Zhu2024mag}\\
			\cline{2-6}~&MEMS &Flip mode&\makecell{Microseconds
				to \\milliseconds order}&\makecell{Micrometer to \\nanometer order}&\cite{MEMS}\\
			\hline
			\multirow{3}*{\makecell{Liquid-\\based\\element}}& Syringe  &1D position                            & \multirow{3}*{\makecell{Milliseconds to \\seconds order}} &\multirow{3}*{Micrometer order}  &\cite{Sytinge}\\
			\cline{2-3}\cline{6-6}~&Nanopump&1D position&  & &\cite{Nano}\\
			\cline{2-3}\cline{6-6}~& Electrowetting&1D position&  & &\cite{Elect}\\
			\hline
			\multirow{3}*{\makecell{Electronically\\reconfigurable\\element}}& Dual-mode patch antenna  &1D position                                 	&\multirow{3}*{\makecell{Nanoseconds to \\milliseconds order}}   &\multirow{3}*{Micrometer order}  &\cite{Ning2025Arch}\\
			\cline{2-3}\cline{6-6}~& PIN diode biasing network&2D rotation  &  & &\cite{PIN}\\
			\cline{2-3}\cline{6-6}~& Pixel antenna&2D position  &  & &\cite{Pixel}\\
			\hline
			\multirow{3}*{\makecell{Movable\\array}} &Wheel gear and motor & Sliding mode      & \multirow{3}*{\makecell{Milliseconds to \\seconds order}}  &\multirow{3}*{Micrometer order}     &\cite{Ning2025Arch}\\
			\cline{2-3}\cline{6-6}~&Motor&Rotatable mode  &  & &\cite{RA1}\\
			\cline{2-3}\cline{6-6}~&Inflatable structure&Foldable mode  &  & &\cite{Inflat}\\
			\hline
		\end{tabular}
	\end{center}
	\vspace{-0.1 in}
\end{table*}

1) \textit{Mechanically Movable Element:} Mechanically movable elements rely on external mechanical structures, which are actuated by devices such as electric motors, precision gears, micro-electromechanical-
systems (MEMS) or motor-driven shafts, to realize physical movement. These actuators transform control signals and energy into mechanical motion, thereby enabling precise positioning and/or rotation of the antenna elements~\cite{Zhu2024mag,Ning2025Arch}. An example of realizing motor-based MA is shown in Fig.~\ref{fig:MA_arch}(a), where the MA is installed on the mechanical slide tracks and driven by step motors. The number of slide tracks and step motors can be flexibly adjusted to enable antenna movement in different spatial DoFs from one-dimensional (1D) to 3D.  Upon receiving the control signal from the central processing unit (CPU), the motors are capable of collaboratively executing the corresponding procedures to reposition the antenna to the target position. Generally, the response time of motor-based MA ranges from milliseconds to seconds with the positioning accuracy on the order of micrometer. In comparison, the MEMS converts energy into micro-scale mechanical motion to drive loads by utilizing physical effects (e.g., deformation, force action) of micro-structures (e.g., piezoelectric, electrostatic, electromagnetic)~\cite{MEMS}. Consequently, the MEMS-based MA can achieve a faster response time on the order of microseconds to milliseconds with a higher positioning accuracy on the order of micrometer or even nanometer. In addition, the power consumption of MEMS-based MA is much less than that of macro motor-based MA, which facilitates the application of MA in energy-constrained scenarios such as mobile and IoT devices. 

2) \textit{Liquid-based Element:} Liquid-based components utilize the flow characteristics of liquid/fluid materials within a container, which can be driven by syringes~\cite{Sytinge}, nanopumps~\cite{Nano}, or electrowetting~~\cite{Elect}. For instance, as shown in Fig.~\ref{fig:MA_arch}(b), liquid metal can move within an air chamber by manually applying pressure to a syringe or digitally controlling a micropump/nanopump. Furthermore, when electrowetting techniques are employed, a voltage applied to the electrodes induces charge redistribution on the surface of the fluid metal. This process alters the surface tension and generates Marangoni forces, which in turn drive the movement of the liquid metal, thereby changing the shape and position of the antenna within the container. In general, the response time of a liquid-based MA is on the order of milliseconds to seconds.

3) \textit{Electronically Reconfigurable Element:} Electronically reconfigurable antennas, such as dual-mode patch antennas~\cite{Ning2025Arch}, are capable of adjusting their phase centers to achieve an equivalent displacement of the active antenna positions. For instance, as shown in Fig.~\ref{fig:MA_arch}(c), by exciting different modes (e.g., TM11 and TM21) in a stacked circular patch antenna, the phase center can be shifted from the physical center, thereby effectively adjusting the antenna position and radiation pattern without involving mechanical motion. Moreover, the PIN diode biasing network represents another efficient approach to realizing rapid reconfiguration of antenna radiation patterns~\cite{PIN}. Additionally, pixel antennas are composed of numerous pixels that can be reconfigured via electronic switches~\cite{Pixel}. Through this mechanism, the position and/or shape characteristics of the antenna are modified, which can be considered an equivalent method for implementing active antenna movement. Owing to their electronic architecture, electronically reconfigurable antennas exhibit a response time ranging from nanoseconds to milliseconds.

4) \textit{Movable Array:} Compared to the element-level movement, the array-level movement sacrifices  part of the spatial DoFs for much less hardware complexity since multiple antenna elements can be moved collectively by using the same drive component. There have been various architectures for implementing array-level movement, such as foldable array~\cite{Inflat}, rotatable array~\cite{RA1,RA2}, and sliding array~\cite{Ning2025Arch}, etc. Specifically, as shown in Fig.~\ref{fig:MA_arch}(d), a foldable array utilizes internal
mechanical structures to possess the capability of expansion and contraction and thus regulates its geometric configurations. Typical methods for achieving foldable arrays include origami-based techniques and inflatable structures. As shown in Fig.~\ref{fig:MA_arch}(e),
a rotatable array is capable of rotational/orientational adjustment with the aid of rotary motors or other means. As shown in Fig.~\ref{fig:MA_arch}(f), a sliding array comprises a single array or multiple sub-arrays, which is capable of sliding along predefined paths or within specific regions (e.g, linear or circular tracks).
The aforementioned  architectures enable a certain degree
of spatial reconfiguration, thereby enhancing wireless transmission
and reception.
For example, the rotations of an array can adjust the main lobe of the antenna radiation pattern to point toward target user clusters, which thus helps improve their effective channel gains
and communication performance. 
\subsection{Channel model}
\label{s2-2}
In this subsection, we discuss the existing channel models for MA, which can be categorized into two branches, i.e., the field-response channel model and the spatial-correlation channel model.
\begin{figure}[t]
	\centering
	\includegraphics[width= 8 cm]{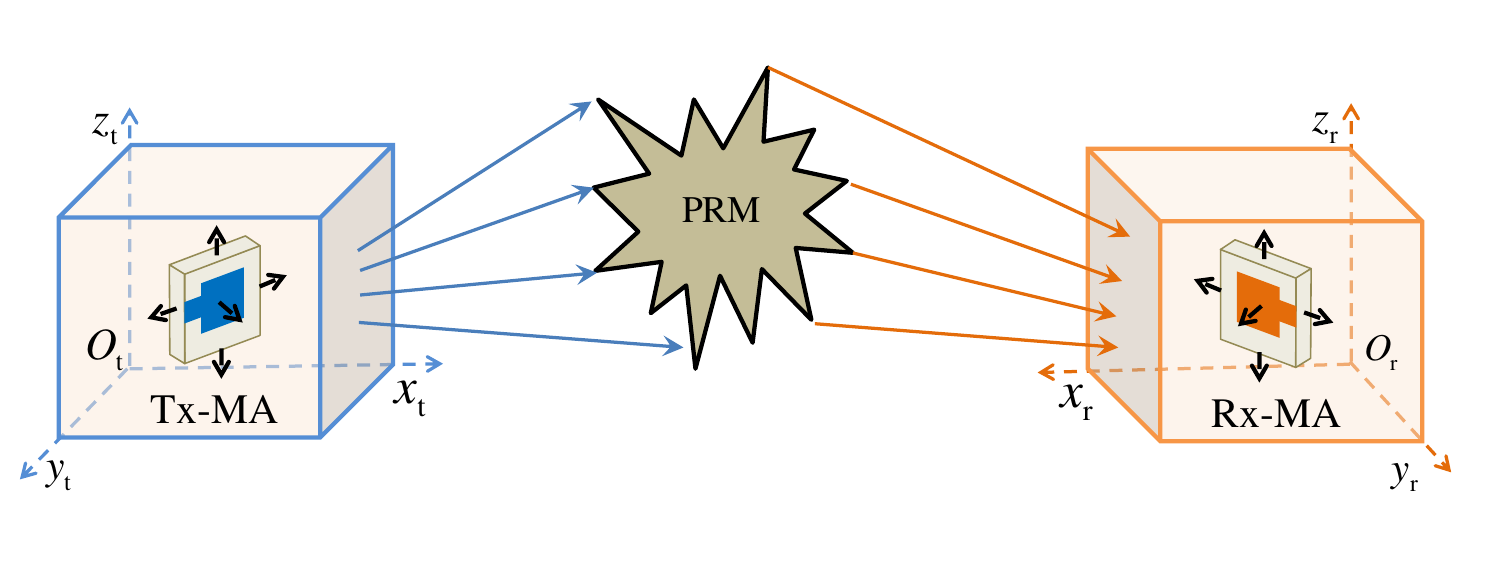}
	\caption{Illustration of the MA-based field-response channel model.}
	\label{fig:FRM}
\end{figure}
\subsubsection{Field-Response Channel Model}
\label{s2-2-1}

We first model the field-response channel  between the Tx-MA and Rx-MA to characterize the channel variation with respect to the MAs' positions. In general, the far-field channel
condition is guaranteed between the Tx and Rx if the
signal propagation distance is much larger than the antenna aperture (i.e, the size of the
moving region in MA systems). For example, for the size of 
moving region $A = 5\lambda$ and the signal wavelength $\lambda = 0.01$ meter (m), the Rayleigh distance can be calculated as $2A^2/\lambda = 0.5$ m. As shown in Fig.~\ref{fig:FRM}, a 3D local coordinate system (LCS) is established to describe the position of the MA at the Tx and Rx, respectively.
The channel response between the transceiver is characterized by
the superposition of coefficients from $L_{\rm t}$ Tx paths and $L_{\rm r}$ Rx paths.
Moreover, under the far-field condition where the planar wave model can be
adopted, the angles of arrival
(AoAs) and angles of departure (AoDs) as well as the amplitudes of the complex path coefficients for multiple channel paths remain unchanged with MAs' movement. Particularly, denote elevation and azimuth AoDs for the $l$-th Tx path as $\theta_{\rm t}^l$ and $\phi_{\rm t}^l$ for $1\le l \le L_{\rm t}$, respectively. Then, the wave vector for the $l$-th Tx path in the Tx-LCS can be obtained as $	\mathbf{k}_{\mathrm t,l}=[\cos\theta_{\rm t}^l\cos\phi_{\rm t}^l,\cos\theta_{\rm t}^l\sin\phi_{\rm t}^l,\sin\theta_{\rm t}^l]^{\mathrm{T}}$. As such, the phase difference in
signal propagation of the $l$-th Tx path between the position of Tx-MA $\mathbf t$ and the reference point (i.e, the origin in the Tx-LCS, $O_{\rm t}$) is expressed as $\mathbf{k}_{\mathrm t,l}^{\rm T}\mathbf{t}$. Accordingly, we can define the field response vector (FRV) to characterize the phase variation of all the Tx channel paths with respect to the position of the Tx-MA as follows~\cite{MASISO_NB}:
\begin{equation}\label{Tx-FRV}
	\mathbf{g}(\mathbf{t})=\left[e^{j\frac{2\pi}{\lambda}\mathbf k_{\mathrm t,1}^{\mathrm T}\mathbf{t}},e^{j\frac{2\pi}{\lambda}\mathbf k_{\mathrm t,2}^{\mathrm T}\mathbf{t}},\dots,e^{j\frac{2\pi}{\lambda}\mathbf k_{\mathrm t,L_{\mathrm t}}^{\mathrm T}\mathbf{t}}\right]^\mathrm{T}.
\end{equation}
Similarly, denote elevation and azimuth AoAs for the $l$-th Rx path as $\theta_{\rm r}^l$ and $\phi_{\rm r}^l$ for $1\le l \le L_{\rm r}$, respectively. Correspondingly, the wave vector for the $l$-th Rx path in the Rx-LCS can be obtained as $\mathbf{k}_{\mathrm r,l}=[\cos\theta_{\rm r}^l\cos\phi_{\rm r}^l,\cos\theta_{\rm r}^l\sin\phi_{\rm r}^l,\sin\theta_{\rm r}^l]^{\mathrm{T}}$. As such, the FRV to characterize the phase variation of all Rx channel paths with respect to the position of the Rx-MA is given by
\begin{equation}\label{Rx-FRV}
	\mathbf{f}(\mathbf{r})=\left[e^{j\frac{2\pi}{\lambda}\mathbf k_{\mathrm r,1}^{\mathrm T}\mathbf{r}},e^{j\frac{2\pi}{\lambda}\mathbf k_{\mathrm r,2}^{\mathrm T}\mathbf{r}},\dots,e^{j\frac{2\pi}{\lambda}\mathbf k_{\mathrm r,L_{\mathrm r}}^{\mathrm T}\mathbf{r}}\right]^\mathrm{T}.
\end{equation}
Moreover, we define the path-response matrix (PRM) as $\mathbf \Sigma \in \mathbb C^{L_{\mathrm t}\times L_{\mathrm r}}$ to characterize the response coefficients between all the Tx and Rx channel paths, where $[\mathbf\Sigma]_{i,j}$ represents the response coefficient between $i$-th Rx and $j$-th Tx channel paths. Notably, the PRM is mainly determined by the antenna characteristics (including orientation, polarization, and radiation pattern) and signal propagation characteristics (including the signal wavelength and propagation environment). However, it remains invariant with changes in the positions of the Tx-MA and Rx-MA, and thus yields the multi-path channel response between the Tx-MA and Rx-MA as follows~\cite{MASISO_NB}:
\begin{equation}\label{MA-channel}
	h(\mathbf{t},\mathbf{r})=\mathbf{f}^{\mathrm H}(\mathbf{r})\mathbf\Sigma\mathbf{g}(\mathbf{t}),
\end{equation}
which is consistent with channel models for conventional FPA systems for the given  positions of Tx-MA and Rx-MA. In particular, the field-response channel model in (\ref{MA-channel}) can be degraded into the specific channel models by restricting the FRVs and PRMs, such as the line-of-sight (LoS) channel (with a single channel path), the geometric channel (with a diagonal PRM). In addition, considering the rich-scatter environment with infinite channel paths, the channel response in (\ref{MA-channel}) can be extended to the integration of path responses over continuous AoAs and AoDs. Nonetheless, for a given moving region size $A$, the angular resolution of antenna can expressed as $\frac{\lambda}{A}$, which indicates the channel paths within an  interval of AoDs/AoAs less than $\frac{\lambda}{A}$ can
be approximately regarded as an effective channel path with an identical AoD/AoA~\cite{tse2005fundamentals}. Consequently, the channel response with infinite channel paths can be approximated by the superposition of responses from finite channel paths, yielding a representation analogous to (\ref{MA-channel}).

The channel model in \eqref{MA-channel} can be easily extended to MA-aided multiple-input single-output (MISO), single-input multiple-output (SIMO), and MIMO systems~\cite{MIMO_FFC,MUMIMO_BS,MUMIMO_user}, as the channel response of each Tx-MA and Rx-MA pair can still be modeled using \eqref{MA-channel}. To maximize the spatial DoFs, larger movable regions are necessary, which may render the far-field plane-wave channel model in \eqref{MA-channel} invalid. In light of this, the near-field spherical-wave model has been developed in~\cite{MIMO_NFC,10909572}, in which the near-field channel response can still be characterized based on antenna positions. To extend from narrow-band flat fading channels to wideband frequency-selective fading channels, a more general multi-tap field-response channel model was proposed in~\cite{MASISO_WB}, in which the wireless channel variations in both space and frequency were characterized by the varying positions of MAs. Overall, these extensions of MA-aided channel models (from single-input single-output (SISO) to MIMO, from far-field to near-field, and from narrow-band to wideband), all express the channel response as a function of  MA positions in an explicit form, laying the performance analysis and optimization in diverse MA-aided wireless systems.

\subsubsection{Spatial-Correlation Channel Model}
\label{s2-2-2}
The spatial-correlation channel model is an alternative way of statistical channel modeling  to simplify the performance analysis and optimization in MA-aided systems. The core idea is to extract the essential statistical characteristics from complicated channel parameters to capture the spatial correlation between wireless channels. Different from the field-response channel model that tracks channel paths, the spatial-correlation channel model directly quantifies statistical dependencies between channel states across different antenna  positions, offering a simplified path to system evaluation.

Early explorations in spatial-correlation channel models leveraged Jake’s model to characterize spatial correlations between discrete antenna positionts. Specifically, under the assumption of Rayleigh fading with rich scattering~\cite{FA1}, a few independent Gaussian random variables are introduced to parameterize channel gains for analytical simplicity. However, this oversimplified structure was latterly proven failing to characterize  correlations across all antenna ports inaccurately~\cite{FA2}.
To improve model accuracy, subsequent work~\cite{FA2} expanded the parameter set to explicitly capture inter-port correlations and introduced some approximations to balance the model precision and analytical tractability. Specifically, for a point-to-point SISO system, an MA at Rx moves freely along
$N$ equally distributed positions (i.e., ports) on a linear space of length $W\lambda$.
Based on Jake’s model, the spatial correlation between the $m$-th and $n$-th ports
is given by~\cite{FA1,FA2}
\begin{equation}\label{Jakes}
	[\mathbf \Lambda_h]_{m,n} = \sigma^2 J_0\left( 2\pi \frac{m - n}{N - 1} W \right),
\end{equation}
for $1\le m,n\le N$, where $\mathbf \Lambda_h\in \mathbb R^{N\times N}$ denotes the spatial correlation matrix of all antenna ports. $\sigma^2$ accounts for the large-scale fading effect and $J_0(\cdot)$
is the zero-order Bessel function of the first kind. As such,
the channel coefficient
at the $n$-th port can be approximately modeled as~\cite{FA2}


\begin{equation}\label{FA-approx}
	\small
	{h}_n = \sqrt{\sigma^2 - \sum_{m=1}^{\epsilon\text{-rank}} u_{n,m}^2 \lambda_m} v_n + \sum_{m=1}^{\epsilon\text{-rank}} u_{n,m} \sqrt{\lambda_m} z_m,
\end{equation}
where $\lambda_1, \lambda_2, \dots, \lambda_{\epsilon\text{-rank}}$  are the non-increasingly ordered eigenvalues of spatial correlation matrix $\mathbf \Lambda_h$ exceeding a threshold $\epsilon$,  with $\epsilon\text{-rank}$ denoting their number. $\mathbf u_1, \mathbf u_2, \dots, \mathbf u_{\epsilon\text{-rank}}$ are their associated eigenvectors
with $\mathbf u_m = [u_{1,m},u_{2,m},\dots,u_{N,m}]^{\text T}$ for $1 \le m \le \epsilon\text{-rank}$. $v_n = c_n + j d_n$ and $z_m = a_m + j b_m$ wherein $c_n, d_n, \forall n$, and $a_m, b_m, \forall m$, are independent and identically distributed
(i.i.d.) Gaussian random variables with zero mean and variance of $\frac{1}{2}$.

Moreover, the  spatial-correlation model was extended to scenarios with 1D continuous antenna movement at the Rx~\cite{FA3} and two-dimensional (2D) discrete movement at both Tx and Rx~\cite{FA4}. In addition, the authors in ~\cite{FA4} proposed a block-fading-based approximation, replacing the full spatial correlation matrix of Rx ports with a block-diagonal structure to simplify performance analysis.

A key advantage of the spatial-correlation channel model is the analytical tractability, which simplifies the derivation of performance metrics such as outage probability and capacity in MA systems. Beyond this, this model exhibits robustness against modeling errors stemming from non-idealities in signal transmission, since a perfect channel structure is difficult to acquire in real-world environments.
Despite these strengths, the framework is constrained by its high reliance on environmental and fading assumptions. Most existing models, such as Jake’s model ~\cite{FA1,FA2,FA3,FA4} and Clarke’s model, depend on predefined conditions such as uniform scattering or Rayleigh fading, limiting their generalizability to different propagation scenarios. Moreover, the channel statistics change with signal propagation and antenna radiation patterns, which cannot be fully characterized by the simple spatial-correlation channel model. Thus,  modeling accurate statistical distributions of wireless channels in different scenarios becomes challenging.

\section{Performance Advantages}
\label{s3}
In this section, we first introduce the typical application scenarios for MA-aided wireless communications and sensing. Then, the performance analysis is conducted to reveal the performance gain of MAs over traditional FPAs.

\subsection{Wireless Communications}
\label{s3-1}
By leveraging the new DoF to fully exploit channel spatial variation, MAs can yield significant performance improvement over traditional FPAs across various communication systems. 

\emph{(1) SISO}: In MA-aided SISO systems, adjusting the position of the antenna enables fine-tuning of the phase shifts across multiple propagation paths. Within the spatial domain, these path coefficients can be constructively superimposed to maximize the overall channel power gain or destructively combined to suppress it. As a result, the antenna position optimization improves communication performance mainly in two ways, i.e., increasing the desired signal power and  reducing interference power. In particular, it has been demonstrated that the MA-SISO system can yield 
significant performance gains over its FPA counterpart for both narrowband and
wideband scenarios~\cite{MASISO_NB,MASISO_WB}, especially for a large number of channel paths.

\emph{(2) MISO/SIMO}: In MISO/SIMO communication systems, MAs enable more effective multi-path channel reconfiguration through joint optimization of their positions. This allows the total channel power gain to be enhanced for desired signal transmission or reduced for interference suppression, which is similar to the behavior observed in SISO systems. Furthermore, when using an MA array, adjusting the positions of multiple antenna elements can dynamically reshape the array geometry. This fundamentally modifies the steering vectors (i.e., array response vectors) of the antenna array. Such reconfiguration makes it possible to adjust the spatial correlations among steering vectors in different directions, thereby supporting more flexible beamforming and advanced array signal processing.

\emph{(3) MIMO}: In MIMO communication systems, the key advantage of  MAs stems from their ability to flexibly reconfigure the MIMO channel matrix, thereby improving spatial multiplexing performance. For instance, in the low-signal to noise ratio (SNR) regime, the channel matrix can be adjusted to maximize its largest singular value, enhancing single-stream beamforming efficiency. Conversely, in the high-SNR regime, optimizing MA positions can help balance the singular values of the channel matrix, better accommodating water-filling based optimal power allocation to facilitate multi-stream transmission. These performance improvements are achievable through MA position optimization, which reconfigures either instantaneous or statistical MIMO channels between the Tx and Rx~\cite{MIMO_FFC,chen2023}.
As revealed in~\cite{MIMO_FFC,MIMO_NFC}, the configuration of antenna positions can change the
effective MIMO channel rank to accommodate different SNR, thereby attaining substantial multiplexing and beamforming gains under both near-field and far-field conditions.

\emph{(4) Multiuser Communications}: The enhanced multiplexing performance offered by MAs can also be leveraged in multiuser communication systems. However, unlike point-to-point MIMO configurations, where both the Tx and Rx use co-located antennas, multiuser systems involve spatially distributed users whose channels are generally independent. In such scenarios, MAs can more effectively reduce inter-user channel correlations, thereby suppressing multiuser interference and achieving greater performance improvements compared to FPA systems. There have been certain works validating that the MAs' position optimization at the user side and/or BS side can effectively improve multiple access channel (MAC) and/or broadcast channel (BC) conditions and suppress multiuser interference~\cite{MUMIMO_BS,MUMIMO_user}, and the statistical channel-based antenna position optimization with much lower movement overhead can also yield a performance comparable to the instantaneous channel-based design~\cite{yan2025}. Furthermore, the extremely large-scale movement of antennas/subarrays (e.g., on the order of several to tens of meters) can proactively establish LoS links between the BS and users as well as decrease inter-user channel correlation, thereby further improving multiuser communication performance~\cite{fu2025}.

\subsection{Wireless Sensing}

\begin{figure}[t]
	\centering
	\includegraphics[width=8 cm]{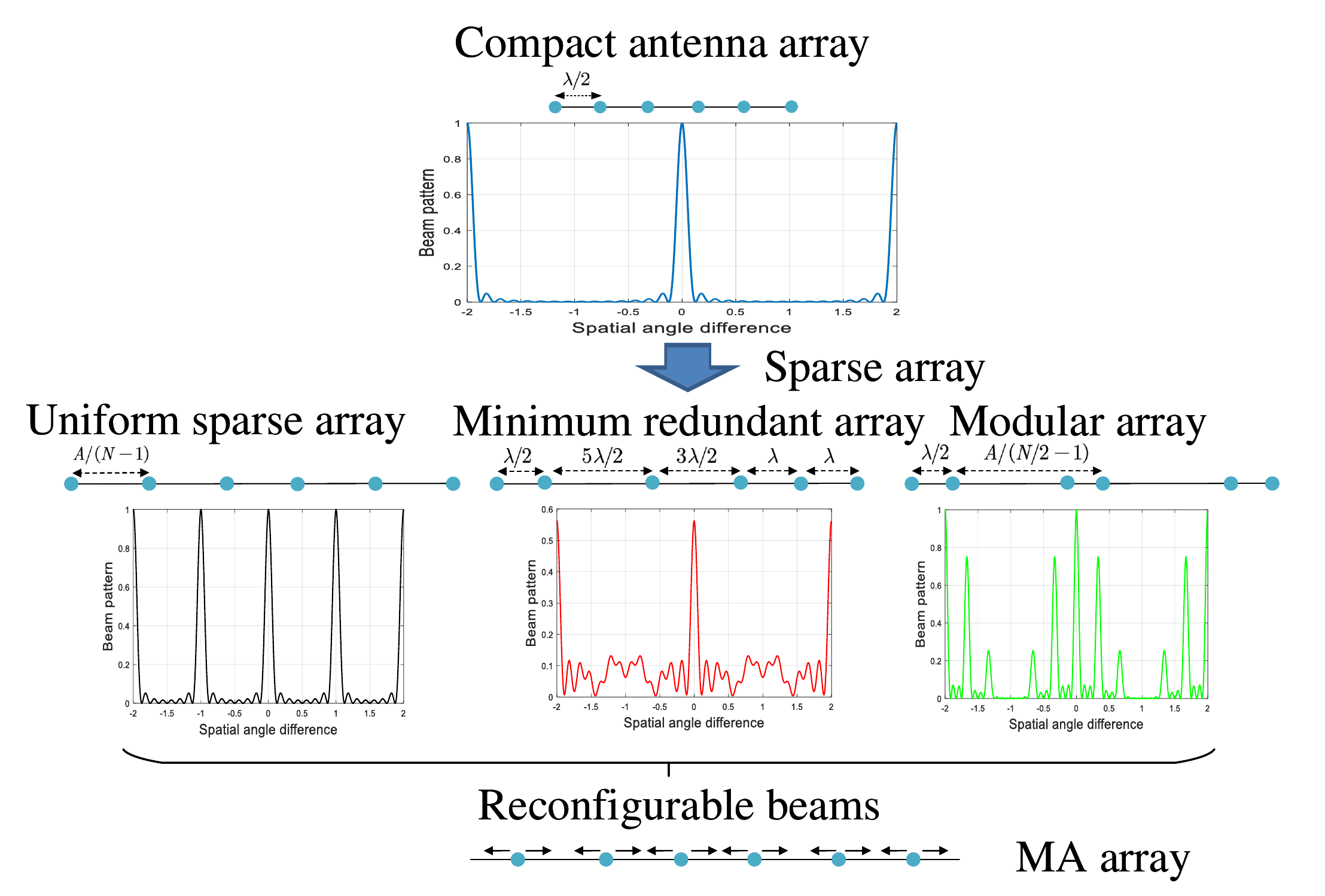}
	\caption{Illustration of MA array-aided wireless sensing systems.}
	\label{fig:MA_array_sensing}
\end{figure}

For existing wireless sensing technology, large-scale compact antenna arrays are usually employed to generate sensing beams with high angular resolution and beamforming gain. However, large-scale antenna arrays may lead to high hardware cost and power consumption. Within this context, multiple forms of sparse arrays, e.g., uniform sparse array, minimum redundant array, and modular array, are developed to reduce the hardware cost while preserving the large aperture. However, the antennas in these sparse arrays are fixed and thus cannot flexibly form the beams to satisfy dynamic wireless sensing requirements. Given the above considerations, the integration of MA technology into  wireless sensing systems can significantly improve the sensing performance by adjusting the geometry of the MA array, as shown in Fig. \ref{fig:MA_array_sensing}. The sensing performance gains include angular resolution improvement, sidelobe suppression, and target decoupling for multi-target sensing. 

\emph{(1) Single-target Sensing}: 
For single-target sensing, the performance gains of the MA-aided sensing system stem from the improvement of precision and accuracy of the target's AoA estimation. Specifically, by enlarging the antenna moving region, the effective array aperture can be increased accordingly, which in turn narrows the main lobe of the sensing beam and thus improves the angular resolution, i.e., the precision of AoA/AoD estimation. Theoretically, it was demonstrated in \cite{10643473,wang2025antenna} that the Cramer-Rao Bound (CRB) of AoA estimation decreases with the increasing variance of antennas' positions. For MA position optimization, the results in \cite{10643473} revealed that the optimized MA positions are concentrated at the edges of the antenna moving region. In addition, the main lobe and side lobes of the receive beam pattern are fundamentally determined by the geometry of the MA array. This indicates that sidelobe suppression can be effectively achieved by optimizing MA positions, thereby rendering AoA estimation robust to interference. For instance, in \cite{10643473}, the beam pattern of the optimized MA array shows a significantly reduced side lobe compared to the conventional uniform sparse array, which in turn significantly improves the AoA estimation accuracy. 

\emph{(2) Multi-target Sensing}: 
In addition to angular resolution improvement and sidelobe suppression, the multi-target sensing also benefits from target decoupling. Mathematically, the derivative of the steering vector (with respect to one target's position) is defined as the sensitivity vector. By adjusting the positions of the antennas, the projection of the sensitivity vector onto the subspace of other users' steering vectors is reduced. This indicates that the angle estimation for one specific target can be less affected by the interference from other targets. In other words, target decoupling allows the multi-target sensing to achieve a comparable performance to sensing individual targets separately, yielding improvements in multi-target sensing performance. In addition to target decoupling, the magnitude of the sensitivity vector (for each target) can be increased by adjusting the positions of the antennas, yielding the improvement of each target's sensing performance. Notably, the two aspects mentioned above may not be achieved simultaneously, which yields a trade-off to obtain the minimum CRB of the AoA estimation for multi-target sensing. 

\begin{figure}[t]
	\centering
	\includegraphics[width= 8 cm]{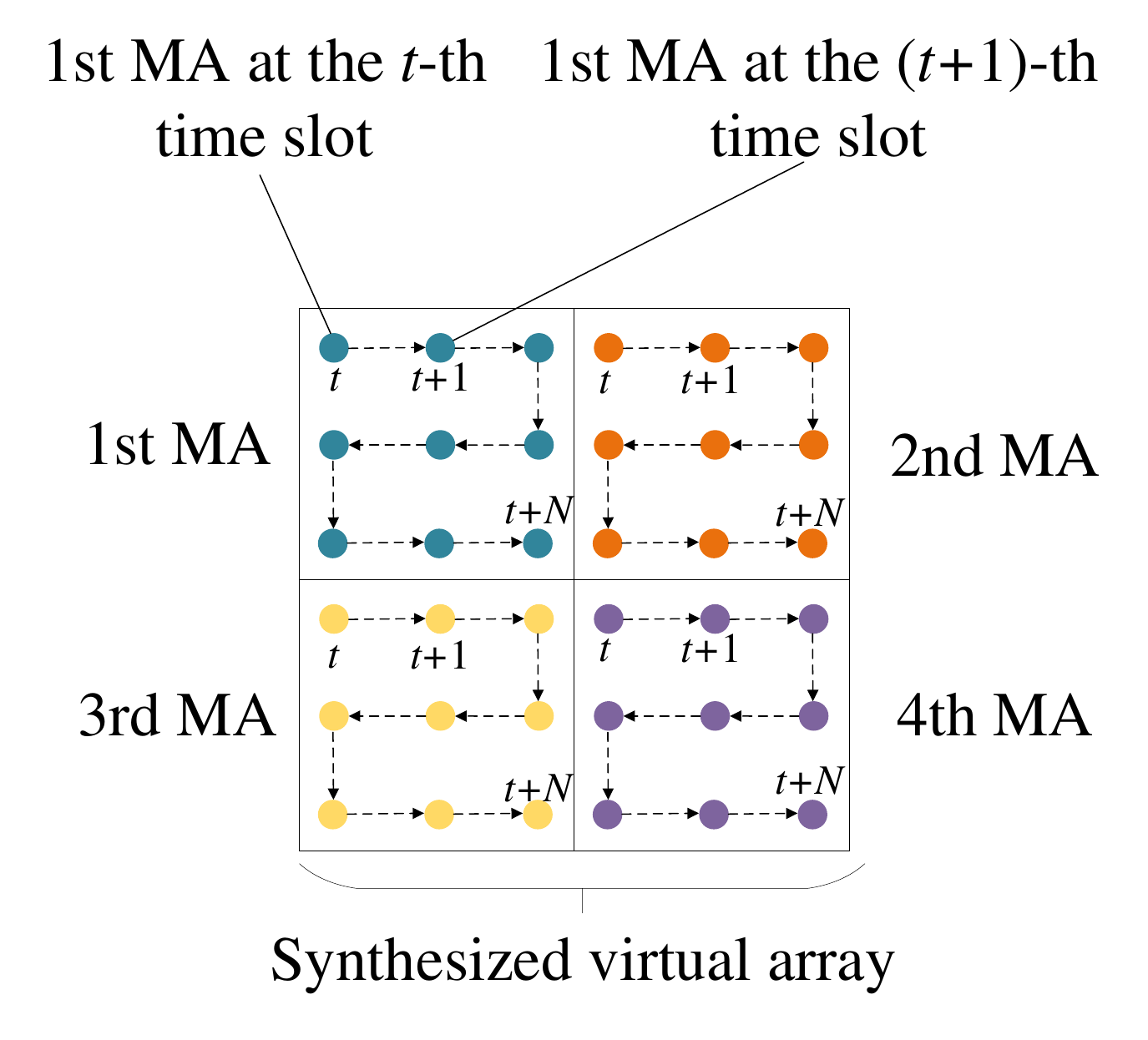}
	\caption{Synthesized virtual array via antenna movement for sensing enhancement.}
	\label{fig:MA_array_time}
\end{figure}

It is worth noting that MA-aided wireless sensing can gain benefits not only from the optimization of MA positions (i.e., the geometry of the MA array) but also from the mobility of the MAs via trajectory optimization~\cite{ma2025}. First, in complex propagation environments with obstacles, the MA-aided wireless sensing system can adjust the positions of the MAs to maintain the LoS link between the sensing nodes and the targets. Second, as shown in Fig. \ref{fig:MA_array_time}, by adjusting the positions of the MAs in real time, one MA can transmit/receive the sensing signal at different positions over multiple consecutive snapshots. Thus, a small number of (even a single) MAs can equivalently synthesize a virtual large-scale antenna array, thus significantly improving the angular resolution and effective DoF for multi-target sensing.

\section{Signal Processing Techniques}
\label{s4}
Although MAs have been shown to offer great potential for enhancing system performance in wireless networks, they introduce new challenges to signal processing techniques due to their distinct operational paradigms compared to traditional FPA systems.  In this section, we investigate the
key technologies for channel acquisition and antenna position optimization in MA-aided wireless networks. 
\subsection{Channel acquisition}
\label{channel_estimation}


The advantages of MA technology have been extensively introduced in Section \ref{s3}. However, the existing studies on MA position optimization highly rely on the knowledge of CSI. In this regard, the channel acquisition serves as a crucial research content for MA system design. 

The channel acquisition for MA systems is more challenging compared to conventional FPA systems. Specifically, for conventional FPA systems, only a limited number of channel responses between discrete positions (where the antennas are located) are required. On the contrary, due to the MAs' ability to move flexibly within the spatial regions, the channel responses between any position pairs within the Tx and Rx regions are required for MA systems. In this regard, conventional channel estimation methods for FPA systems cannot be directly applied to MA systems. 

\subsubsection{Narrowband Channel Estimation}\label{MA_narrow_CE}


Several researches on the acquisition of instantaneous CSI for MA-aided narrowband communication systems have been carried out. These methods can be divided into three categories, i.e., model-based, model-free, and artificial intelligence (AI). In the following, each of these three categories will be presented.

\emph{(1) Model-based  Approach}: Based on the field-response channel model, the model-based channel acquisition methods estimate the AoDs, AoAs, and complex gains of the multi-path components (MPCs) to calculate the channel responses. Specifically, compared to the infinite number of channel responses in the locational domain, the number of the MPCs is finite, leading to a finite number of AoDs, AoAs, and complex gains. This indicates that the MPC information can be recovered with a limited number of pilots. Then, with the estimated MPC information, the channel response between any position pairs within the Tx region and Rx region can be calculated by the field-response channel model accordingly.

To estimate the MPC information,  existing schemes include the compressed sensing-based method \cite{10497534,10236898} and the tensor decomposition-based method \cite{10659325}. By discretizing the angular domain, the compressed sensing-based method formulates the estimation of the AoDs and AoAs as a sparse signal recovery problem, which is then solved by compressed sensing algorithms. Subsequently, complex gains are estimated via the least-square (LS) method. Specifically, we take the joint MPC information estimation method in \cite{10497534} as an example, in which an MA-SISO system with 2D Tx and Rx moving regions is considered. The coordinates of the Tx- and Rx-MAs are ${\bf t} = \left[x_{\rm t}, y_{\rm t}\right]^{\rm T}\in\mathbb{C}^{2\times 1}$ and ${\bf r} = \left[x_{\rm r}, y_{\rm r}\right]^{\rm T}\in\mathbb{C}^{2\times 1}$, respectively. Define virtual elevation and azimuth AoDs as $\vartheta_{\rm t}^{l} = \cos\theta_{\rm t}^{l}\cos\phi_{\rm t}^{l}$ and $\varphi_{\rm t}^{l} = \cos\theta_{\rm t}^{l}\sin\phi_{\rm t}^{l}$ for the $l$-th Tx path, $l = 1,\cdots, L_{\rm t}$, respectively. Similarly, define the virtual elevation and azimuth AoAs as $\vartheta_{\rm r}^{l} = \cos\theta_{\rm r}^{l}\cos\phi_{\rm r}^{l}$ and $\varphi_{\rm r}^{l} = \cos\theta_{\rm r}^{l}\sin\phi_{\rm r}^{l}$ for the $l$-th Rx path, $l = 1,\cdots, L_{\rm r}$, respectively. Then, discrete virtual AoD/AoA sets are defined by quantizing $\left[-1,1\right]$ into $G$ grids to approximate any virtual AoDs/AoAs (which range from $-1$ to $1$), i.e., 
\begin{equation}
	\begin{aligned}
		&\tilde{\bf \Theta}_{v} = \left\{\tilde{\vartheta}_{v}^{g_{vx}} = -1+\frac{2g_{vx}-1}{G}\left| g_{vx} = 1,\cdots,G\right.\right\},\\
		&\tilde{\bf \Phi}_{v} = \left\{\tilde{\varphi}_{v}^{g_{vy}} = -1+\frac{2g_{vy}-1}{G}\left| g_{vy} = 1,\cdots,G\right.\right\},
	\end{aligned}
\end{equation}
in which $v\in\left\{{\rm t,r}\right\}$. With the discrete virtual angles, based on \eqref{Tx-FRV} and \eqref{Rx-FRV}, discrete FRVs can be defined as
\begin{equation}
	\begin{small}
		\begin{aligned}
			&\tilde{\mathbf{g}}\left({\bf t}\right) = \left[ e^{j\frac{2\pi}{\lambda} y_{\rm t} \tilde{\vartheta}_{\rm t}^{g_{ty}}} \right]_{1 \leq g_{ty} \leq N} \otimes \left[ e^{j\frac{2\pi}{\lambda} x_{\rm t} \tilde{\varphi}_{\rm t}^{g_{tx}}} \right]_{1 \leq g_{tx} \leq N},\\
			&\tilde{\mathbf{f}}\left({\bf r}\right) = \left[ e^{j\frac{2\pi}{\lambda} y_{\rm r} \tilde{\vartheta}_{\rm r}^{g_{ry}}} \right]_{1 \leq g_{ry} \leq N} \otimes \left[ e^{j\frac{2\pi}{\lambda} x_{\rm r} \tilde{\varphi}_{\rm r}^{g_{rx}}} \right]_{1 \leq g_{rx} \leq N}.
		\end{aligned}
	\end{small}
\end{equation}
Similarly, discrete PRM is defined as $\tilde{\bf \Sigma}\in\mathbb{C}^{N^2\times N^2}$, containing all the response coefficients with the corresponding discrete virtual angles. Then, based on \eqref{MA-channel}, the discrete field-response channel model is given by 
\begin{equation}\label{eq:discrete_field_response}
	\begin{aligned}
		h\left({\bf t,r}\right)& = \tilde{\bf f}\left({\bf r}\right)^{\rm H}\tilde{\bf \Sigma}\tilde{\bf g}\left({\bf t}\right)+e\left({\bf t,r}\right)\\
		&\overset{(a)}{=}\left[\tilde{\bf g}\left({\bf t}\right)^{\rm T}\otimes \tilde{\bf f}\left({\bf r}\right)^{\rm H}\right]{\bf u}+e\left({\bf t,r}\right),
	\end{aligned}
\end{equation}
in which $e\left({\bf t,r}\right)$ and ${\bf u}\in\mathbb{C}^{N^4\times 1}$ represent the quantization error and vectorized PRM, respectively, and step (a) represents the vectorization of discrete PRM $\tilde{\bf \Sigma}$. With the discrete field-response channel model, the received signal $v$ can be represented as
\begin{equation}
	y = \sqrt{P}\left\{\left[\tilde{\bf g}\left({\bf t}\right)^{\rm T}\otimes \tilde{\bf f}\left({\bf r}\right)^{\rm H}\right]{\bf u}+e\left({\bf t,r}\right)\right\}s + n,
\end{equation}
in which $P$, $s$, and $n$ represent the transmit power, transmit signal, and complex additive Gaussian noise, respectively. For channel estimation, the Tx- and Rx-MAs simultaneously move over $M$ positions for channel measurements. Without loss of generality, we let $s = 1$ during the channel measurements. The $m$-th positions of the Tx- and Rx-MAs are denoted as ${\bf t}_{m} = \left[x_{\rm t}^{m},y_{\rm t}^{m}\right]^{\rm T}$ and ${\bf r} = \left[x_{\rm r}^{m}, y_{\rm r}^{m}\right]^{\rm T}$, $m = 1,\cdots,M$. Then, the received pilots ${\bf y}\in\mathbb{C}^{M\times 1}$ is given by
\begin{equation}\label{eq:pilot_signal}
	\begin{aligned}
		\mathbf{y} &= \sqrt{P}\left\{ \begin{bmatrix}
			\tilde{\mathbf{g}}\left({\bf t}_{1}\right)^{\rm T} \otimes \tilde{\mathbf{f}}\left({\bf r}_{1}\right)^{\rm H} \\
			\vdots \\
			\tilde{\mathbf{g}}\left({\bf t}_{M}\right)^{\rm T} \otimes \tilde{\mathbf{f}}\left({\bf r}_{M}\right)^{\rm H}
		\end{bmatrix} \mathbf{u} + \mathbf{e} \right\} + \mathbf{n} \\
		&\triangleq \sqrt{P} \boldsymbol{\Psi} \mathbf{u} + \sqrt{P} \mathbf{e} + \mathbf{n},
	\end{aligned}
\end{equation}
where ${\bf e}\in\mathbb{C}^{M\times 1}$ and ${\bf n}\in\mathbb{C}^{M\times 1}$ represent the quantization error and noise vectors over $M$ positions. Matrix ${\bf \Psi}\in\mathbb{C}^{M\times N^4}$ is defined as the measurement matrix. Then, with \eqref{eq:pilot_signal}, the MPC information can be estimated via the following sparse signal recovery problem
\begin{equation}\label{eq:mpc_recovery}
	\begin{aligned}
		\min_{\mathbf{u}} &\quad \|\mathbf{u}\|_0, \\
		\text{s.t.} &\quad \|\mathbf{y} - \sqrt{P} \boldsymbol{\Psi} \mathbf{u}\|_2 \leq \|\mathbf{y}\|_2 \epsilon_0,
	\end{aligned}
\end{equation}
where $\epsilon_0$ is the parameter to guarantee the minimization of the channel estimation error. Problem \eqref{eq:mpc_recovery} can be solved via compressed sensing methods to estimate vectorized PRM ${\bf u}$, which contains the MPC information, and the channel response between arbitrary position pairs of the Tx- and Rx-MAs can be recovered via the discrete field-response channel model, i.e., \eqref{eq:discrete_field_response}. Notably, for each estimated entry in ${\bf u}$, its index uniquely determines the elevation/azimuth AoDs and AoAs of a channel path, while the complex gain is determined by the value of this entry. However, the joint estimation of the MPC information may lead to a high computational complexity. In such a case, the authors in \cite{10236898} proposed a successive transmitter-receiver compressed sensing (STRCS) method, in which the AoDs, AoAs, and complex gains are estimated successively. In particular, in the first step, the Tx-MA moves over different positions for channel measurements while the Rx-MA is fixed. With the received pilots, a sparse signal recovery problem can also be formulated similar to problem \eqref{eq:mpc_recovery}, in which the AoDs of all the Tx paths can be estimated via the compressed sensing method. It is worth noting that in the first step, only the AoDs can be estimated, which is different from the joint MPC estimation method. Similarly, in the second step, the Tx-MA is fixed while the Rx-MA moves over different positions for channel measurements. Similarly, the AoAs can also be recovered by the compressed sensing method based on the formulated sparse signal recovery problem. Next, in the third step, the Tx-MA and Rx-MA move simultaneously for additional channel measurements. With the channel measurements in the three steps and the estimated angles, the complex gains are recovered via the LS method. It is worth noting that the STRCS method reduces the dimension of the measurement matrix in the compressed sensing method, leading to a significantly reduced computational complexity. 
However, the existing compressed sensing-based channel estimation methods are on-grid methods, i.e., angular discretization is required. These methods inherently suffer from resolution limitations. In other words, an insufficient number of angular discretization grids may reduce the accuracy of the estimated AoDs and AoAs, which in turn severely degrades the channel estimation performance. On the other hand, a sufficiently large number of grids may lead to a high computational complexity.  

The tensor decomposition-based channel estimation method achieves off-grid estimation of the AoDs and AoAs by exploiting the Vandermonde structure of the factor matrices (obtained from the canonical polyadic (CP) decomposition of the received pilots)~\cite{10659325}. Specifically, under the MIMO setup, one Tx-MA moves over different positions while all the Rx-MAs remain unchanged in the first step. Then, the CP decomposition is utilized to obtain the Tx factor matrices, which inherently exhibit the Vandermonde structure for AoD estimation. In the second step, all the Tx-MAs are fixed while the Rx-MAs move over different positions. Similarly, with the received pilots in the second step, the factor matrices can be obtained via the CP decomposition, and the AoAs can be estimated by exploiting the Vandermonde structure accordingly. Finally, with the estimated AoDs and AoAs, the complex gains can be estimated via the LS method. For the tensor decomposition-based channel estimation method, the off-grid estimation significantly increases the accuracy of the MPC information and thereby improves the channel estimation performance. However, this method confines MA measurement positions to a rectangular structure, thus limiting the design flexibility.

\emph{(2) Model-free Approach:} The model-free channel estimation performs channel measurements at a small number of discrete positions, and then estimate the channel response at other unmeasured positions via interpolation. It can be developed based on strong spatial correlation \cite{9992289,10278813}, Fourier transform \cite{10751774}, and Bayesian linear regression \cite{zhang2023successive,cui2024near}. In \cite{9992289,10278813}, for an arbitrary unmeasured position, its channel response was assumed to be the same as that of the nearby position. This method requires the MA measurement positions to be sufficiently dense to guarantee the accuracy of channel estimation, which inevitably leads to extremely high pilot overhead. In \cite{10751774}, the Fourier transform was first performed on the received to obtain the channel response in the frequency domain. Then, a low-pass filter was utilized to remove the impact of the limited region size and discrete MA measurement positions. Finally, the CSI of the entire region was reconstructed by performing the inverse Fourier transform on the frequency-domain channel response. The Bayesian linear regression methods regard the channel response in the confined region as a random process with respect to the positions of the MAs, in which each channel realization (i.e., instantaneous CSI) is regarded as a sample of the random process \cite{zhang2023successive,cui2024near}. Based on this framework, the channel response of the entire region can be obtained by weighting the channel responses at the MA measurement positions. Specifically, the Bayesian linear regression methods contain two stages, i.e., offline design and online regression. In the offline design stage, the MA measurement positions and the weighting matrix are determined by the kernel, which characterizes the spatial correlation of the channel. In the online regression stage, the MAs move over the pre-defined MA measurement positions. Then, based on the channel responses and the weight matrix, the channel response of the entire region is reconstructed with low complexity, since only matrix multiplication is performed. However, it is worth noting that the above studies on the model-free approaches only estimate the channel responses at discrete positions, and fail to capture the continuous channel variation in the spatial domain. Moreover, due to the lack of specific channel structures, the model-free approaches may require extremely high pilot overhead for larger-size and higher-dimensional antenna moving regions.


\emph{(3) AI-based Approach}: The AI-based channel estimation methods require the measured channel responses as the input to predict the channel responses at the unmeasured positions. In general, the AI-based methods contain two stages, i.e, the offline training and the online deployment. In the offline training stage, the data set with channel responses at both measured and unmeasured positions is utilized to train the network. In the online deployment stage, by taking the measured channel responses as input to the network, the network can predict the channel responses at the unmeasured positions. This indicates that only a low computational complexity is required in the online deployment stage for channel prediction. However, it is worth noting that there are AI-based channel measurement results based on a real propagation environment for MA systems  in the literature, which can serve as the training set for AI-based methods. This highlights the importance of the channel measurement campaign for MA systems. 

The AI-based channel estimation methods can be either model-free \cite{tang2025accurate,10447499,10495003,huang2025cnn} or model-based \cite{xu2024sparse,how@Kang,jang2025new}. For model-free channel estimation methods, the prediction of the channel responses is mainly achieved based on the spatial correlation. For instance, in \cite{tang2025accurate}, a diffusion model (DM) with a simplified U-Net architecture was trained to capture the spatial correlation of the channel. Then, in the online deployment stage, the CSI of the entire region was reconstructed via a denoising diffusion restoration model (DDRM) while the DM is applied as an implicit prior. In \cite{10447499}, a ResNet-based network with a hard selection approach was proposed, which contained three sub-networks to be dynamically selected based on the spatial correlation conditions to balance the channel estimation performance and pilot overhead. By leveraging the local spatial correlation and smoothness of the channel in the confined region, an asymmetric graph masked autoencoder (AGMAE), which incorporates the attention mechanism as well as the local diffusion mechanism of graph neural networks is proposed for CSI reconstruction \cite{10495003}. In addition to the 2D region, the study on reconstructing the CSI of the 3D region was conducted in \cite{huang2025cnn}, in which the convolutional neural network (CNN) was utilized.

In addition to the model-free methods for predicting channel responses based on the spatial correlation, model-based channel estimation can also be implemented using AI-based methods. In \cite{xu2024sparse}, the channel estimation problem was transformed as a basis selection problem, which was then solved by the proposed sparse Bayesian learning (SBL)-based channel estimator. Moreover, the MA measurement positions were optimized by maximizing the mutual information increment between adjacent received pilots. To further reduce the pilot overhead and computational complexity, the latent domain representation of the channel was obtained through the eigenvalue decomposition (EVD) of the spatial correlation matrix, which was then recovered by jointly optimizing the training overhead, training sequences, and discrete MA measurement position switching \cite{how@Kang}. Moreover, it reveals that the training overhead for estimating the channel at all the discrete positions is always less than the rank of the spatial correlation matrix of all the discrete positions. However, the AI-based channel estimation methods mainly focus on the recovery of the channel responses at discrete positions. In \cite{jang2025new}, a novel model-based AI-based channel estimation method was proposed. Instead of reconstructing the CSI directly, the trained neural network predicts the AoDs and AoAs. Then, the complex gains and the corresponding reconstructed CSI are obtained via the gradient-based alternating minimization algorithm in the following refinement stage. With the MPC information, the channel response at an arbitrary position in the continuous region can be reconstructed. 

The channel estimation methods for MA systems are summarized in Table \ref{tab:ce}. The model-based methods focus on the estimation of MPC information and channel reconstruction via the field-response model, thus yielding a high accuracy for the channels with few paths. Moreover, for large antenna moving regions, a relatively smaller number of  channel measurements are required since the channel exhibits sparsity in the angular domain compared to the channel responses in the spatial domain. However, the model-based methods rely on accurate channel models and may be sensitive to channel measurement errors. In contrast, the model-free methods aim to reconstruct the channel responses directly from the measured channel responses. This indicates that specific channel structures are not required, and thus these methods are robust to the increase in the number of paths. However, on one hand, for large antenna moving regions, extremely high pilot overhead is required to guarantee the accuracy of channel reconstruction. On the other hand, existing studies on the model-free methods only focus on the estimation of channel responses at discrete positions at the Rx side only, which thus limit the DoFs for antenna movement. The AI-based methods focus on predicting the channel responses using the trained models, leading to a low computational complexity for the online deployment stage.
\begin{table*}[t]
	\centering
	\caption{A Summary of Channel Estimation Methods for MA Systems.}
	\label{tab:ce}
	\begin{tabular}{|m{1.5cm}|m{4cm}|m{4cm}|m{4.8cm}|m{1.2 cm}|}
		\hline
		& \makecell{\textbf{Advantages}} & \makecell{\textbf{Limitations}} & \makecell{\textbf{Estimation Method}} &\makecell{\textbf{Ref.}} \\
		\hline
		\multirow{3}{=}[-4ex]{Model-based approaches} & 
		\multirow{3}{=}[-1ex]{1) High accuracy for channels with few paths; \\2) Lower channel measurement requirements for large moving regions.} & 
		\multirow{3}{=}[-2ex]{1) Rely on accurate channel models; \\2) Sensitive to channel measurement errors.}  & \makecell{Joint MPC \\information estimation\\ via compressed sensing} & \cite{10236898} \\
		\cline{4-5}
		&  &   &
		\makecell{Successive MPC \\information estimation \\via compressed sensing}& 
		\cite{10497534} \\
		\cline{4-5}
		&  &  & \makecell{Successive MPC \\information estimation \\via tensor decomposition}  & \cite{10659325}\\
		\hline
		\multirow{5}{=}{Model-free approaches} & 
		\multirow{5}{=}[-1.5ex]{1) Specific channel models are not required; \\ 2) Robust to the increase in the number of paths.} &
		\multirow{5}{=}[-1ex]{ 1) Extremely high pilot overhead for large moving regions; \\ 2) Only estimate the channel responses at discrete positions.} & \makecell{Channel responses at unmeasured \\ positions are assumed to \\be the same as that at\\ nearby measured positions} &  \cite{9992289,10278813}\\
		\cline{4-5}
		&  &  & \makecell{Channel response estimation \\ via Fourier transform} & \cite{10751774}\\
		\cline{4-5}
		&  &  & \makecell{Channel response \\estimation via Bayesian \\ linear regression} & \cite{zhang2023successive,cui2024near}\\
		\hline
		\multirow{6}{=}{AI-based approaches} & 
		\multirow{6}{=}{} & 
		\multirow{6}{=}{} & \makecell{Channel response estimation \\via model-free \\ AI-based methods} & \cite{tang2025accurate,10447499,10495003,huang2025cnn} \\
		\cline{4-5}
		& {Low computational complexity requirement for the online deployment stage.} & High computational complexity for offline training.  & \makecell{Channel response estimation \\via model-based \\ AI-based methods} & \cite{xu2024sparse,how@Kang}\\
		\cline{4-5}
		&  &  & \makecell{Successive MPC information \\ estimation  via model-based \\ AI-based method} & \cite{jang2025new}\\
		\hline
	\end{tabular}
\end{table*}

\subsubsection{Extension}

As shown in Section \ref{MA_narrow_CE}, there have been numerous studies focusing on MA-aided narrowband instantaneous channel estimation in point-to-point systems. Research on aspects such as MA-aided wideband channel estimation, multiuser channel estimation, and statistical channel estimation is still at a preliminary stage. 

\emph{(1) Wideband Channel Estimation}: 
Existing MA-aided wideband channel estimation methods are mainly model-based \cite{cao2025channel,dong2025group}. For instance, in \cite{cao2025channel}, the field-response channel model is extended to wideband systems. Then, similarly to the STRCS method in \cite{10236898}, by performing channel measurements at different positions, the AoDs and AoAs can be estimated successively via the compressed sensing method. For wideband systems, by placing the pilots across different subcarriers, the delays can be estimated accordingly. With the recovered angles and delays, the complex gains can be recovered by the LS method. Moreover, an alternating refinement method is also proposed, in which the AoDs, AoAs, and delays are further refined via the off-grid gradient descent method by minimizing the discrepancy between the received pilots and those constructed by the estimated MPC information. Instead of estimating the MPC information, the channel frequency responses are directly recovered via the compressed sensing-based method in \cite{dong2025group}. Specifically, this paper first maps the channel frequency responses to the wavenumber-delay domain, which exhibits the group sparsity due to the finite antenna aperture and bandwidth. Then, by leveraging the group sparsity, the channel frequency response estimation problem is formulated as a sparse signal recovery problem and solved via the compressed sensing methods.  

\emph{(2) Statistical Channel Estimation}: Although most existing studies  focus on reconstructing instantaneous CSI of the entire MA moving region, such CSI may vary over time due to the change of propagation environment, especially for high-mobility scenarios. In other words, the reconstructed CSI may fail to reflect the current channel conditions, preventing the communication system from achieving sufficient performance gains after MA position optimization, which is based on the reconstructed instantaneous CSI. In such a case, existing studies have explored the MA position optimization based on statistical CSI, e.g., \cite{yan2025, chen2023,zheng2025two}, which may remain unchanged for a long period. However, the acquisition of statistical CSI for MA systems still remains unexplored. The statistical CSI can be regarded as a random process, while the instantaneous CSI is one sample of the process. To reconstruct the statistical CSI, one may measure the instantaneous CSI over a long period offline. Then, the statistical CSI can be obtained by averaging the obtained samples. However, this method requires substantial time overhead for channel measurements. On the other hand, the statistical channel estimation methods for conventional FPA systems, in which the instantaneous CSI is not required, may be extended to MA systems to reduce the time overhead. For instance, the angular domain channel power matrices acquisition method in \cite{10146318} may be extended to estimate the distribution of the AoDs and AoAs for statistical CSI acquisition.

\emph{(3) Near-field Channel Estimation}: With the increase of antenna moving region and carrier frequency, the Tx and Rx may be located within the Rayleigh distance of each other. Thus, the near-field field-response channel model based on the spherical wave model should be adopted~\cite{10909572}. In this regard, the corresponding near-field channel estimation methods for MA systems are required. On one hand, the model-based channel estimation methods may be able to extend to near-field channel estimation under LoS-dominant or path-sparse scenarios. One possible way is to successively estimate the LoS path and the NLoS paths \cite{10078317}. Specifically, with the received pilots, the angle, distance, and complex gain of the LoS path are estimated via the gradient descent method by minimizing the discrepancy between the received pilots and those constructed by the estimated MPC information of the LoS path. Then, based on the received pilots without the effect of LoS path, the estimation of the MPC information for NLoS paths can also be formulated as a sparse signal recovery problem, and can be solved via the compressed sensing methods to obtain the AoDs, AoAs, distances, and complex gains from the scatterer to the center of Tx-MA and Rx-MA moving regions of the NLoS paths. On the other hand, model-free channel estimation methods may be applied to near-field scenarios, while the spatial correlation should be distinguished with that for far-field channels. 

\emph{(4) Multiuser Channel Estimation}: Most existing MA channel estimation methods focus on point-to-point systems. However, channel estimation for MA-aided multiuser systems may introduce new challenges, since the channels between the BS and all the users are required. One possible way is that the BS transmits common pilots to the users, and the users perform channel estimation simultaneously. Then, the users feed back the reconstructed CSI to the BS. For instance, in \cite{10375559}, the downlink scenario was considered, in which the BS transmitted the common pilots to the users, and the users estimated the MPC information via the compressed sensing method simultaneously. For uplink scenarios, one possible way is to adopt orthogonal time/frequency resource blocks for pilot transmission and reception. However, this method may lead to high pilot overhead when the number of users is large. Thus, the method for channel estimation in MA-aided multiuser systems remains a significant gap in current research, and further exploration is still required.
\begin{figure}[t]
	\centering
	\includegraphics[width= 8 cm]{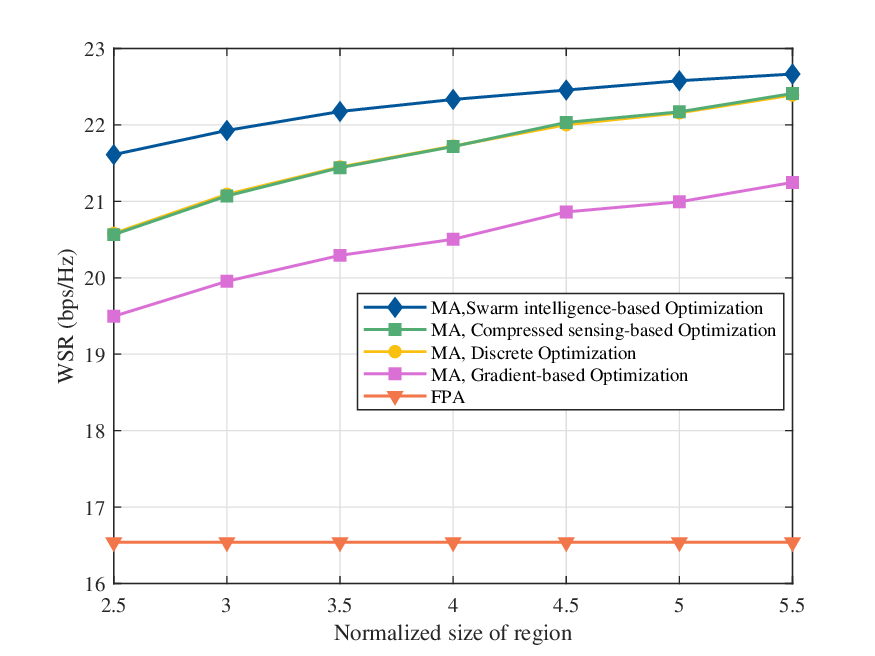}
	\caption{Performance comparison of different antenna position optimization methods.}
	\label{fig:performance}
\end{figure}
\subsection{Antenna Position Optimization}
\label{s4-2}
Note that the aforementioned application scenarios attain performance improvement owing to the deployment of MAs at the optimized positions. 
However, due to the highly nonlinear relationship between antenna positions and  channel responses, the antenna position optimization faces huge challenges in balancing between the  computation  complexity and optimization performance. To address this, significant research efforts have been devoted to developing efficient antenna position optimization algorithms, which can be categorized into three types, i.e, CSI-based, CSI-free, and AI-empowered methods, as summarized in Table~\ref{tab:APV}.

\begin{table*}[h]
	\renewcommand{\arraystretch}{1.3}
	\caption{Comparison of methods for antenna position optimization.}\label{tab:APV}
	\footnotesize
	\begin{center}
		\begin{tabular}{|c|c|c|c|c|c|}
			\hline
			\textbf{}
			&\textbf{Methods}           &\textbf{\makecell[c]{Limitations}} &\textbf{\makecell[c]{Optimization\\Performance}}    &\textbf{\makecell[c]{Computational\\Complexity}} &\textbf{Ref.}\\
			\hline
			\multirow{5}{*}[-20pt]{\makecell[c]{\vspace{0pt}CSI-\\based\\approaches}}& Closed-form solution              & \makecell{Only suitable for \\simple and specific cases}  &Optimal &\makecell{Proportional
				to number of \\ MAs}&\cite{MASISO_NB,MANull_steering}\\
			\cline{2-6}
			& Gradient-based Optimization& \makecell{Likely falling into \\a local optimum}&Suboptimal &\makecell{Proportional
				to numbers \\of iterations and MAs}&\cite{MIMO_FFC,MUMIMO_user}\\
			\cline{2-6}
			& Swarm intelligence-based Optimization& \makecell{High computational \\complexity} &Near-optimal &\makecell{Proportional
				to numbers of \\MAs, swarm size, \\and moving region size}&\cite{MUMIMO_BS,GAMA}\\
			\cline{2-6}
			& Discrete Optimization& \makecell{Sacrifice part of \\spatial DoFs} &Suboptimal &\makecell{Proportional to numbers of \\MAs and candidate positions}&\cite{GraphMA,BnBMA}\\
			\cline{2-6}
			& Compressed sensing-based Optimization
			& \makecell{Constrained by sparse \\signal recovery framework} &Suboptimal &\makecell{Proportional to numbers of \\MAs and candidate positions}&\cite{MIMO_NFC,CS1}\\
			\hline
			\multirow{2}{*}{\makecell{CSI-\\free\\approaches}}&ZO gradient gradient approximation& \makecell{High online \\training overhead} &Non-optimal &\makecell{Proportional to numbers of \\MAs and measured positions} &\cite{zorder}\\
			\cline{2-6}
			& Bayesian optimization & \makecell{High online \\training overhead} &Non-optimal &\makecell{Proportional to numbers of \\MAs and measured positions}&\cite{BO}\\
			\hline
			\multirow{2}{*}{\makecell{AI-\\empowered\\approaches}}& CSI-based RL/DL & Prohibitive training overhead &Near-optimal &\makecell{Proportional	to numbers of \\MAs and network parameters} &\cite{DL1,RL1,DL2}\\
			\cline{2-6}
			& CSI-free RL/DL & Challenging training task  &Suboptimal &\makecell{Proportional
				to numbers of \\MAs and network parameters}&\cite{Zhu2024mag}\\
			\hline
		\end{tabular}
	\end{center}
	\vspace{-0.1 in}
\end{table*}

\emph{(1) CSI-based Approach}: Based on the perfectly known CSI, the globally optimal solutions
for MA positions can be derived in closed form in some simple cases~\cite{MASISO_NB,MANull_steering}, where the system setups and optimization objectives are generally specific and simplified, such as single MA or pure LoS channels. Although the closed-form solutions ensure the best system performance with minimal computational complexity, they cannot be applied to most practical communication systems with multi-antenna/multi-user/multi-path setups. To address this, locally optimal search techniques are widely explored in the related works.

The gradient-based ascent/descent optimization updates the antenna position along//opposite the gradient direction of the objective function. In general, this method yields a suboptimal solution with a relatively low computational complexity~\cite{MIMO_FFC,MUMIMO_user}. However, in cases where the objective function is highly non-convex/non-concave with respect to the antenna position, directly applying gradient-based optimization in the non-convex optimization problem may fall into an undesired local optimum. To cope with this challenge, the successive convex approximation (SCA) technique is commonly employed for relaxing the optimization problem and transforming it from non-convex/non-concave  to convex/concave. Nevertheless, the slack introduced to the original problem will inevitably lead to optimization performance degradation, and how to minimize such performance degradation becomes crucial for the design of gradient-based optimization.

In comparison, swarm intelligence-based optimization offers stronger global search capabilities, at the cost of higher computational complexities. Examples include particle swarm optimization (PSO)~\cite{MUMIMO_BS,PSOzhang}, the genetic algorithm~\cite{GAMA}, and hippopotamus optimization~\cite{NOMA2024Xiao}, which optimize the antenna position by exploiting individual and swarm experiences to fully explore the entire antenna moving region. Notably, the  performance of swarm intelligence optimization is mainly determined by the swarm size and the number of iterations, which, however, will lead to prohibitive computational complexity with the increasing number of antennas  in the large movable region in XL-MIMO. Consequently, there is a trade-off between system performance and computational complexity in swarm intelligence-based optimization, and it is essential to adaptively adjust the optimization settings to accommodate the requirements of different scenarios.

In addition, by sacrificing part of the spatial DoFs in antenna movement to reduce the dimension of the solution space, continuous position optimization can be converted into discrete versions via sampling of the antenna moving region. Examples include graph-based solutions~\cite{GraphMA} and branch-and-bound (BnB) methods~\cite{BnBMA}, which select the optimal antenna position from the discrete position set. However, as the moving region size or sampling resolution increases, the computational complexity of these approaches grows exponentially in the worst-case scenario.

Recently, compressed sensing-based methods have emerged as low-complexity solutions for antenna position optimization~\cite{MIMO_NFC,CS1}. The key idea is to transform the original optimization problem into a sparse signal recovery problem, where only a small number of positions corresponding to the activated MAs are selected from a large feasible region. It can be efficiently solved by existing techniques, such as orthogonal matching pursuit	(OMP). Compared to the above solutions, this approach achieves a favorable trade-off between global search capability and computational complexity. Nevertheless, not all application scenarios can accommodate a sparse signal recovery framework, as it requires specific system configurations and design objectives to be compatible with it.

To validate the effectiveness of the CSI-based methods for antenna position optimization, we consider a multiuser MISO downlink communication system with the BS equipped with an MA array and each user equipped with a single FPA. As shown in Fig.~\ref{fig:performance}, we compare the maximum weighted sum-rate (WSR) metric for different methods
versus the normalized size of the antenna moving region. It can
be seen that all the MA schemes are superior to the FPA schemes, which demonstrates the effectiveness  of antenna position optimization. 
The swarm intelligence-based scheme is superior to all the other schemes at the cost of the highest computational complexity. Moreover, compared to the discrete and compressed sensing-based optimizations, 
the gradient-based method suffers from performance degradation since it generally
falls into a local optimum. Overall, it is suggested to select appropriate methods
based on different optimization objectives and constraints  to achieve the desired trade-off between optimization performance and computational complexity.

\emph{(2) CSI-free Approach}: Most of the above
works assume accurate CSI for antenna position optimization, which requires a large number of channel estimations and leads to additional pilot overhead, especially in fast-varying propagation channels.
To cope with these challenges, CSI-free optimization has emerged as an effective method that circumvents explicit for channel parameters and thus truns to be robust against CSI errors. The basic idea is to input the performance measurements (e.g., channel gain, SNR, achievable rate) at  a small number of measured positions to a black-box optimizer, and finally output the optimized antenna positions. Compared to the CSI-based methods, the CSI-free optimization requires a much smaller number of channel measurements and pilots at the cost of acceptable performance degradation. Examples include zeroth-order (ZO) gradient approximation~\cite{zorder} and Bayesian optimization. Specifically, in~\cite{zorder}, the antenna position optimization problem was first transformed into a derivative-free problem, where the gradient of the objective function was approximately calculated by using ZO gradient approximation techniques. Then, based on ZO adaptive momentum method (AdaMM) algorithm, the antenna position was adaptively updated and finally converges to a desired  local optimum. In~\cite{BO}, a training-based position optimization approach was proposed that operates without explicit channel estimation. Specifically, during pilot transmission, the received signal strength was directly used to determine the antenna position for each iteration. To improve search efficiency, BO was employed to balance exploration-exploitation trade-offs, thereby minimizing pilot overhead.

\emph{(3) AI-empowered  Approach}: In recent years, the advancement of AI technology has provided a robust tool for addressing complex and non-convex optimization problems, thereby efficiently determining the optimal solutions for antenna position and resource scheduling. Examples include deep learning (DL), reinforcement learning (RL), and deep reinforcement learning (DRL). Compared to the sophisticated iterative algorithms widely employed in traditional optimization methods, the offline training and online deployment strategy in AI-based schemes significantly reduces the computational complexity while achieving superior performance, thereby facilitating real-time and low-cost wireless communication.

Similar to traditional optimization methods, AI-empowered approaches can be categorized into two classes based on whether CSI is known or not. For CSI-based DL, channel parameters are directly employed as inputs to the neural network, whereas antenna positions serve as outputs of the network. For instance, in~\cite{DL1}, a DL model was designed for multicast beamforming with an MA array. The DL model comprised three modules: a feature extractor, an antenna position vector (APV) optimizer, and an antenna weighting vector (AWV) optimizer, with each module constructed using a feedforward neural network (FNN). By formulating the objective function as the loss function, the three modules were jointly trained in an unsupervised manner, ultimately outputting the optimized APV and AWV.

For CSI-based RL, channel parameters and antenna positions are typically employed to represent the state and action of the agent for antenna movement, respectively. For instance, in~\cite{RL1}, a multi-agent deep deterministic policy gradient (MADDPG) algorithm was proposed for enhancing the capacity of the MA-enabled multi-receiver communication system. In particular, the RL model comprised two types of agents: beamforming agents and MA agents, which were employed to learn strategies for beamforming and antenna movement. The strategies of multiple agents were collaboratively updated based on rewards obtained through interaction with the environment. Based on offline training under various imperfect CSI conditions, MADDPG can ultimately output solutions for transmit beamforming and antenna movement in real time.

CSI-free based AI techniques remain an open research problem requiring further investigation. Nevertheless, the solution can be inspired by traditional CSI-free designs. Specifically, the CSI-free optimizer aims to establish a mapping from a finite set of channel measurements to optimized antenna positions. Thus, motivated by the robust capability of neural networks in fitting non-linear relationships,  offline training can be conducted via gradient backpropagation of the loss function, which implicitly incorporates the statistical characteristics of wireless channels. Then, without high-complexity channel estimation, by directly inputting pilot signals received at a moderate number of positions into the well-trained neural networks, optimized antenna positions are outputted, thereby improving system performance as guaranteed by the loss function.

Although the aforementioned methods for antenna position optimization are primarily tailored for wireless communication systems, they can also be extended to wireless sensing systems~\cite{10643473,SensingMA}, as their solution concepts share commonalities. For example, an MA array-aided wireless sensing system was studied in~\cite{10643473} to improve angle estimation accuracy through antenna position optimization. Specifically, a globally optimal solution for MAs’ positions was derived in closed form for the case of a 1D MA array, while a locally optimal solution for MAs’ positions was achieved via an efficient alternating optimization algorithm for the case of a 2D MA array. In~\cite{SensingMA}, a block coordinate descent (BCD) algorithm with momentum gradient descent and logarithmic barrier penalties was proposed to jointly optimize transmit and receive MAs’ positions. It was designed to maximize the weighted sum of CRBs, thereby improving sensing performance. Overall, the methods mentioned in Table~\ref{tab:APV} can be flexibly selected for antenna position optimization in wireless sensing system designs based on specific system configurations and design objectives.

%

\section{Future Direction}
\label{s5}
In this section, we discuss several promising application scenarios for MA technology as well as the corresponding challenges that still remain unaddressed, aiming to inspire further studies in this area. 

\subsection{Large-scale Antenna Movement}
Most studies on MAs have mainly concentrated on wavelength-scale movement regions, where variations are confined to phase shifts. This inherent limitation results in a restricted capacity to mitigate large-scale path loss. To address this constraint, the large-scale antenna movement has recently been studied, which allows for flexible movement of antennas/subarrays over
an extremely large region and thus facilitates higher DoFs in antenna placement~\cite{fu2025,LSMA}. An alternative realization of large-scale antenna movement is
the pinching antenna technique~\cite{PA1,PA2}, which activates antennas by attaching small dielectric particles (e.g., plastic pinches) to a dielectric waveguide. The flexible positioning of these pinching antennas along an extended length enables them to move closer to target users, facilitating the establishment of strong LoS channels and reducing large-scale path loss.

However, pinching antennas equipped with a shared RF chain can only operate on a preconfigured 1D waveguide. This restriction limits both their movement flexibility and spatial multiplexing performance. Furthermore, most existing studies on pinching antenna systems have adopted the assumption of pure LoS channels, which may not accurately represent real-world propagation environments characterized by multipath effects.
Moreover, how to implement large-scale antenna movement in an extremely large 2D/3D region faces certain challenges in response time and power overhead, which needs more research efforts, especially in hardware structure design.

\subsection{Hybrid movable and reconfigurable antennas}
Different from the  MAs that focus on the  adjust external placement (including position and orientation)~\cite{MASISO_NB,shao20256DMA}, the reconfigurable antennas (RAs) have gained attraction for their ability to reconfigure internal radiation characteristics, including radiation pattern, polarization, and frequency response~\cite{Recon1,Recon2}. 
Despite their differences, both MAs and RAs share the
unique capability of reconfiguring wireless channels directly
in the electromagnetic domain. Consequently, the hybrid movable and reconfigurable antennas can combine the advantages of both while alleviating their inherent limitations to flexibly adjust
antenna properties in higher DoFs, thereby improving  spectrum and energy efficiency~\cite{zhu2025movable}.
For instance, electronic RA elements can be embedded into a mechanical MA array, which delivers enhanced flexibility for reconfiguring channel characteristics. Specifically, large-scale positional adjustments can be achieved by manipulating the mechanical array, enabling adaptation to slowly varying statistical channels. In contrast, small-scale adjustments to antenna position or radiation characteristics can be realized through tuning the electronic elements, allowing the system to adapt to rapidly varying instantaneous channel conditions.

However, most of the exiting research on wireless systems enabled by MAs and RAs has been conducted independently, with the performance benefits of their integrated design remaining unexplored. In this scenario, a unified channel model that captures the effects of both antenna movement and electromagnetic configuration is   essential, since it serves as the foundation for enabling joint optimization. Furthermore, deriving network performance bounds based on this unified model is also critical to unlock their anticipated potential for enhancing wireless communication and/or sensing performance. However, the existing literature on this specific topic remains relatively limited.

\subsection{MA-enhanced ISAC}

Integrated sensing and communications (ISAC), which merges wireless sensing and communication into one single system, has become a key technology for the upcoming 6G networks. By enabling the ability to share the hardware and spectrum, ISAC technology simultaneously achieves  sensing and communication with significant cost reduction. The MA technology can bring new potentials to ISAC systems. First, the real-time adjustment of the MAs' positions enables the MA-aided ISAC systems to meet diverse requirements when switching between the communication and sensing tasks. Second, the trade-offs between the communication performance and sensing accuracy can be balanced by optimizing the positions of the MAs. Third, by optimizing the geometry of the MA array, the correlation between the steering vectors across different directions can be reduced, thereby decreasing angle estimation ambiguity for wireless sensing and mitigating interference for multiuser communication. Fourth, the movement of the antennas enlarges the array aperture, providing higher angle estimation resolution for wireless sensing and improving spatial multiplexing performance for wireless communication. However, in practical scenarios, the frequent changes of the sensing targets' and communication users' locations challenge the real-time adjustment of the MA positions. Thus, the optimization of the MA positions based on statistical channels or user/target distributions is a promising direction for MA-aided ISAC systems. Moreover, existing studies on MA-aided ISAC systems focus on the sensing for a single target. Thus, the MA position optimization under the scenarios with multiple targets under clutters is still an open issue.

\subsection{MA-empowered Space-Air-Ground-Sea Integrated Network}
Space-air-ground-sea integrated network has been proposed to support the diverse services and demands for various scenarios. In this context, MAs can be deployed on various platforms, serving a promising technology to enhance the communication performance in the space-air-ground-sea integrated network. Specifically, for satellite communication, high directional beams are usually required to compensate the significant path losses. Additionally, low earth orbit (LEO) satellites require to serve the ground users with distinct distributions and coverage requirements. In this context, the adjustment of the MA array geometry can reduce the correlation between the steering vectors over the coverage and interference directions, achieving flexible coverage. Moreover, due to the high mobility of the LEO satellites, the coverage and interference areas as well as the beam directions may change over time, and thus flexible beamforming is required. The real-time adjustment of the MA array geometry enables the optimization of the beams over time. Notably, the limited power may pose new constraints and challenges for MA-aided satellite communication systems. For instance, the geometry of the MA array may not be able to change frequently due to the limited power, requiring a dedicated study on the MA geometry optimization based on slowly-varying statistical channel knowledge for satellite communication.

With the increase of the unmanned aerial vehicles (UAVs) payload capacity and minimization of communication devices, UAVs can also be integrated into the space-air-ground-sea integrated network as the aerials, relays, and users. The flexible 3D deployment of the UAV, enables them to avoid the blockages in scenarios with dense obstacles, such as urban areas. In this context, by deploying the MAs on UAVs, the small-scale movement of the MAs can cooperate with the large-scale movement of the UAVs for communication performance improvement. Specifically, the UAVs can move to positions with fewer obstacles for better channel condition, while the movement of the MAs can reduce the channel correlation between different users and thereby mitigate interference. Despite the potentials, MA-aided UAV communication poses new challenges. For instance, the UAV wobbling caused by wind disturbance leads to inaccurate antenna positions, degrading the communication performance of MA-aided UAV communication systems. Thus, the MA position optimization design that is robust to inaccurate MA positions requires further research. 

With the development of maritime activities, such as offshore oil exploration and naval shipping, the integration of maritime users into a space-air-ground-sea integrated network has attracted attention from both academia and industry. The channel characteristics of maritime communications exhibit significant differences compared to conventional terrestrial communications. Due to the sparsity of the scatterers above the sea surface, the LoS path and reflection path (reflected by the sea surface) are dominant. Moreover, the fluctuations of sea waves and the mobility of maritime communication users introduce additional randomness into the channel. The design of MA systems in maritime environments should take these unique channel characteristics into consideration such that the potential can be fully unleashed.

\section{Conclusion}
\label{s6}
In this paper, we provided a comprehensive overview
of MA-enhanced wireless networks. The integration of MA into wireless communication and sensing has catalyzed a fundamental paradigm shift in wireless system design, transitioning from conventional FPA architectures to intelligent frameworks featuring reconfigurable and dynamically adaptive antenna arrays. First, we introduced the basic models of MAs, including the antenna movement model and the channel model. Whereafter,
the performance advantages of MAs across different  application scenarios were also demonstrated. Then, we discussed the signal processing techniques for channel acquisition and antenna position optimization. Finally, several promising
open problems and  promising research directions were illuminated to inspire further investigation in this area.

\bibliographystyle{cjereport}
\bibliography{refs}

\end{document}